\documentclass[preprint]{aastex}

\usepackage[latin1]{inputenc}
\usepackage{graphicx}
\usepackage[english]{babel}
\usepackage[english]{layout}
\usepackage{fullpage}
\usepackage{array}
\usepackage{amsmath}
\usepackage{textcomp}
\usepackage{amssymb}
\usepackage{amsfonts}
\usepackage{color}
\usepackage{here}
\usepackage{braket}
\usepackage{multicol}
\usepackage{natbib}
\usepackage{pdfpages}
\usepackage{ulem}
\usepackage{multicol}
\usepackage{float}

\renewcommand{\thesection}{\arabic{section}}
\renewcommand{\thesubsection}{\arabic{section}.\arabic{subsection}}
\newcommand{\dd}{^\circ~{\rm day}^{-1}}
\newcommand{\dpi}{\dot\varpi}
\newcommand{\ares}{a_{\rm res}}

\begin{document}

\title{\huge{How Janus' Orbital Swap Affects \\ the Edge of Saturn's A Ring?}}

\author{\Large{{\noindent Maryame El Moutamid$^{*1}$}, 
{Philip D. Nicholson$^1$}, 
{Richard G. French$^2$},
{Matthew S. Tiscareno$^3$},
{Carl D. Murray$^4$}, 
{Michael W. Evans$^1$}, 
{Colleen McGhee French$^2$},  
{Matthew M. Hedman$^5$}, 
and {Joseph A. Burns$^1$}}\\ 
\vspace{0.5cm}
\small{$^1$ { Department of Astronomy, Cornell University, Ithaca, NY 14853, USA}\\
$^2$ { Department of Astronomy, Wellesley College, Wellesley, MA 02481, USA}\\
$^3$ { SETI Institute, Mountain View, CA 94043, USA}\\
$^4$ { Astronomy unit, School of Physics and Astronomy, Queen Mary University, London, UK}\\
$^5$ { Department of Physics, University of Idaho, Moscow, ID 83844-0903, USA}}}

\maketitle

\noindent $^*$ Corresponding author : maryame@astro.cornell.edu

\bigskip
{\bf ABSTRACT} 
\bigskip

We present a study of the behavior of Saturn's A ring outer edge, using images and occultation data obtained by the Cassini spacecraft over a period of 8 years from 2006 to 2014.  More than 5000 images and 170 occultations of the A ring outer edge are analyzed.  Our fits confirm the expected response to the Janus 7:6 Inner Lindblad resonance (ILR) between 2006 and 2010, when Janus was on the inner leg of its regular orbit swap with Epimetheus.  During this period, the edge exhibits a regular 7-lobed pattern with an amplitude of 12.8 km and one minimum aligned with the orbital longitude of Janus, as has been found by previous investigators.  However, between 2010 and 2014, the Janus/Epimetheus orbit swap moves the Janus 7:6 LR away from the A ring outer edge, and the 7-lobed pattern disappears.  
In addition to several smaller-amplitudes modes, indeed, we found a variety of pattern speeds with different azimuthal wave numbers, and many of them may arise from resonant cavities between the ILR and the ring edge; 
also we found some other signatures consistent with tesseral resonances that could be associated with inhomogeneities in Saturn's gravity field. 
Moreover, these signatures do not have a fixed pattern speed.
We present an analysis of these data and suggest a possible dynamical model for the behavior of the A ring's outer edge after 2010.



\section{\bf Introduction}

\begin{figure}[!h]
\centerline{\includegraphics[angle=0,totalheight=9cm,trim= 0 0 0 0]{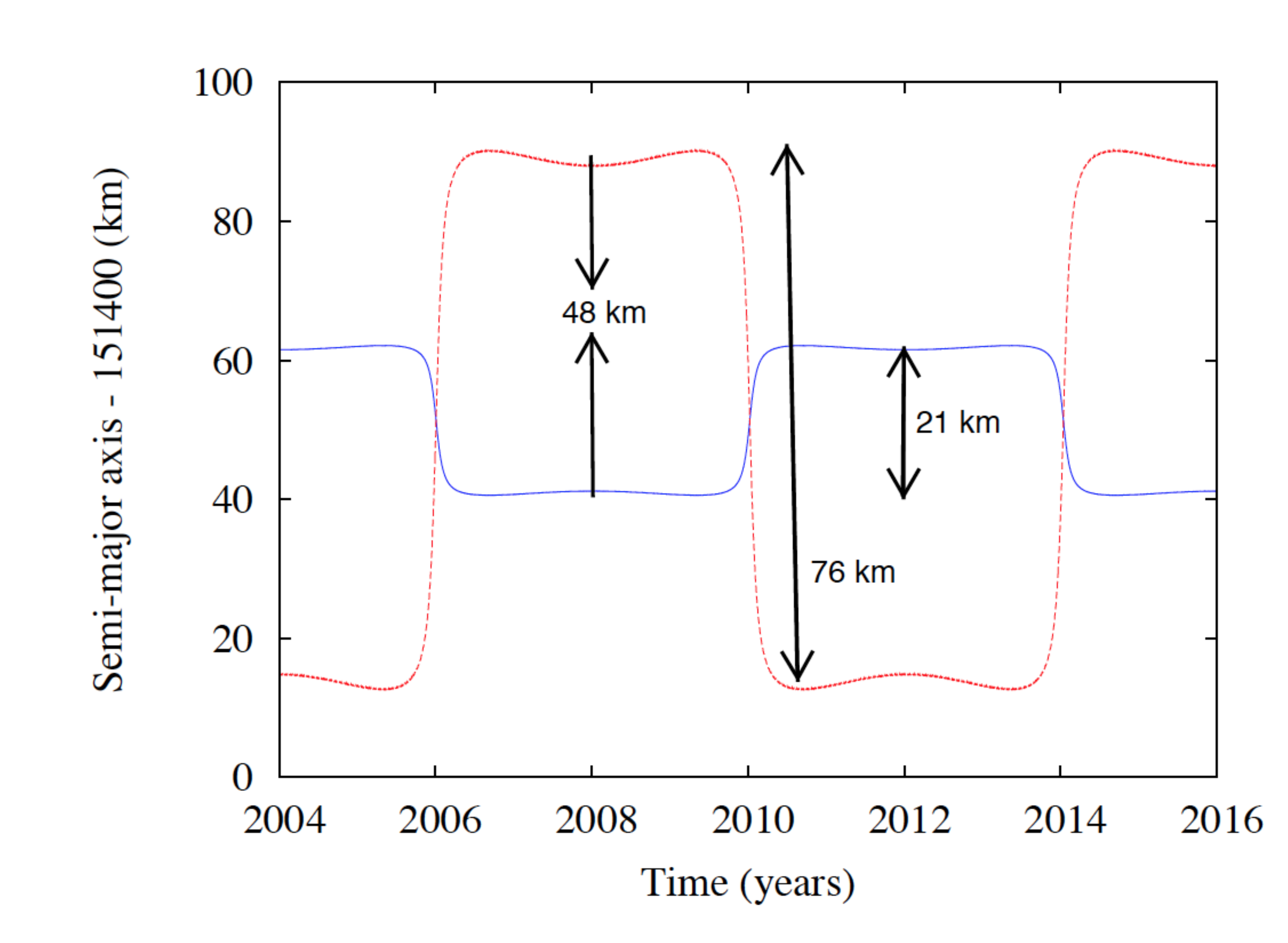}} 
\caption{This figure displays the orbital semimajor axes  --- computed from the epicyclic theory \citep{renner06} --- of Saturn's small moons Janus (blue) and Epimetheus (red) over 12 years. The two moons occupy very nearly the same mean orbit and swap orbital positions relative to Saturn once every four years, with Janus moving by  $\pm21$~km and Epimetheus by  $\pm76$~km. Their mean orbital separation is  $\sim48$~km, less than the physical radius of either body. The long-term mean semi-major axis is $151,451.6$~km \citep{jacobson08}.}
\label{orbit_evol}
\end{figure}

The saturnian satellite system is endowed with an unusually large number of orbital resonances \citep{peale86}. 
Among these are several small satellites in corotation resonances like Anthe, Methone and Aegaeon \citep{elmoutamid14}, and Trojan satellites that are in 1:1 resonance with the medium-sized moons Tethys and Dione \citep{murray05,robutel12}, as well as the pair of co-orbital satellites, Janus and Epimetheus.  The latter are unique in the solar system, and represent an extreme form of 1:1 resonance where the two bodies are similar in mass.  Both objects share a common mean orbit, with a period of $\sim16.7$~hours, but at any given instant their semimajor axes differ by $\sim48$~km (see Fig. \ref{orbit_evol}) and their mean motions by $0.25\dd$.  Their mean radii are $89.2 \pm 0.8$ and $58.2 \pm 1.2$~km,  respectively \citep{thomas13}.

Every four years, the two moons approach one another closely, exchange angular momentum via their mutual gravitational attraction and then recede again, having switched their relative orbital radii \citep{murray99}. 
At each orbital exchange, Janus' semimajor axis shifts inwards or outwards by $\sim21$~km while that of Epimetheus shifts in the opposite direction by $\sim76$~km, reflecting their mass ratio of 3.6.
Their overall configuration repeats itself every 8.0 years.  
Their relative motion is analogous to the horseshoe libration of a test particle in the restricted three body problem around the Lagrange points $L_4$ and $L_5$. In a reference frame rotating at their long-term average mean motion,
Janus moves in a small, bean-shaped orbit, while the less massive Epimetheus moves in a more elongated horseshoe-type orbit. The reader is referred to \cite{yoder83} or \cite{murray99} for a more detailed description of the dynamics involved, supported by numerical integrations.
The relative libration amplitudes of the two moons combined with their observed libration period make it possible to determine their individual  masses quite accurately, despite their relatively small sizes \citep{yoder89,nicholson92}. 

Current estimates for the mean orbital elements, masses and radii of the coorbital moons are based on observations by the \textit{Cassini} spacecraft, combined with earlier \textit{Voyager} and ground-based measurements \citep{jacobson08}. In Fig.~\ref{orbit_evol} we illustrate the predicted variations in both satellites' semimajor axes over a period of 12 years, based on a numerical integration with initial conditions obtained for
epoch 2004 from the JPL Horizons on-line database.  This diagram illustrates two aspects of their motion that should be kept in mind: (1) neither satellite's semimajor axis is precisely constant even when the two bodies are far apart, and (2) the actual orbital exchanges are not instantaneous, but occur over a period
of a few months, or quite slowly relative to orbital timescales. 

Many of Saturn's satellites generate observable features within the planet's extensive  ring system, either in the form of spiral density waves driven at mean motion resonances or at the sharp outer edges of several rings \citep{tiscareno07,colwell09}. 
In addition to driving several strong density waves, as well as numerous weaker waves \citep{tiscareno06}, the coorbital satellites appear to control the location and shape of the outer edge of the A ring.  Based on a study of Voyager imaging and occultation data, \cite{porco84} concluded not only that this edge is located within a few km of the Janus/Epimetheus 7:6 ILR, but that it exhibits a substantial 7-lobed radial perturbation that rotates around the planet at the same rate as the satellites' average mean motion.  The situation appeared to be quite analogous to the control of the outer edge of the B ring by the Mimas 2:1 ILR,
also studied by  \cite{porco84}.  But the limited quantity of high-resolution Voyager data, combined with the fact that the Voyager 1 and 2 encounters were separated by only 9 months, meant that it was not possible to examine the effect of the coorbital swap on the ring edge.\footnote{The Voyager encounters occurred in Nov 1980 and Aug 1981, while the nearest orbital swaps occurred in January 1978 and January 1982. At the time of both encounters, Janus was the outer satellite.}

The Cassini mission, launched in 1997, arrived at the Saturn system in July 2004 and the spacecraft has been taking data continuously for over 10 years. In this time, it has been able to
observe the effects of the coorbital satellites'  orbital exchanges in January 2006, January 2010 and in January 2014.  During this period, over 30 sequences of 
images covering most of the circumference of the A ring edge have been obtained, to study both the F ring and the outer edge of the A ring, and over 150 stellar and radio occultation experiments have been performed. 
This dataset provides the first opportunity to study the behavior of the A ring edge throughout the complete 8-year coorbital cycle, including both configurations of the Janus-Epimetheus system.

\begin{figure}[!h]
\centerline{\includegraphics[angle=0,totalheight=9cm,trim= 0 0 0 0]{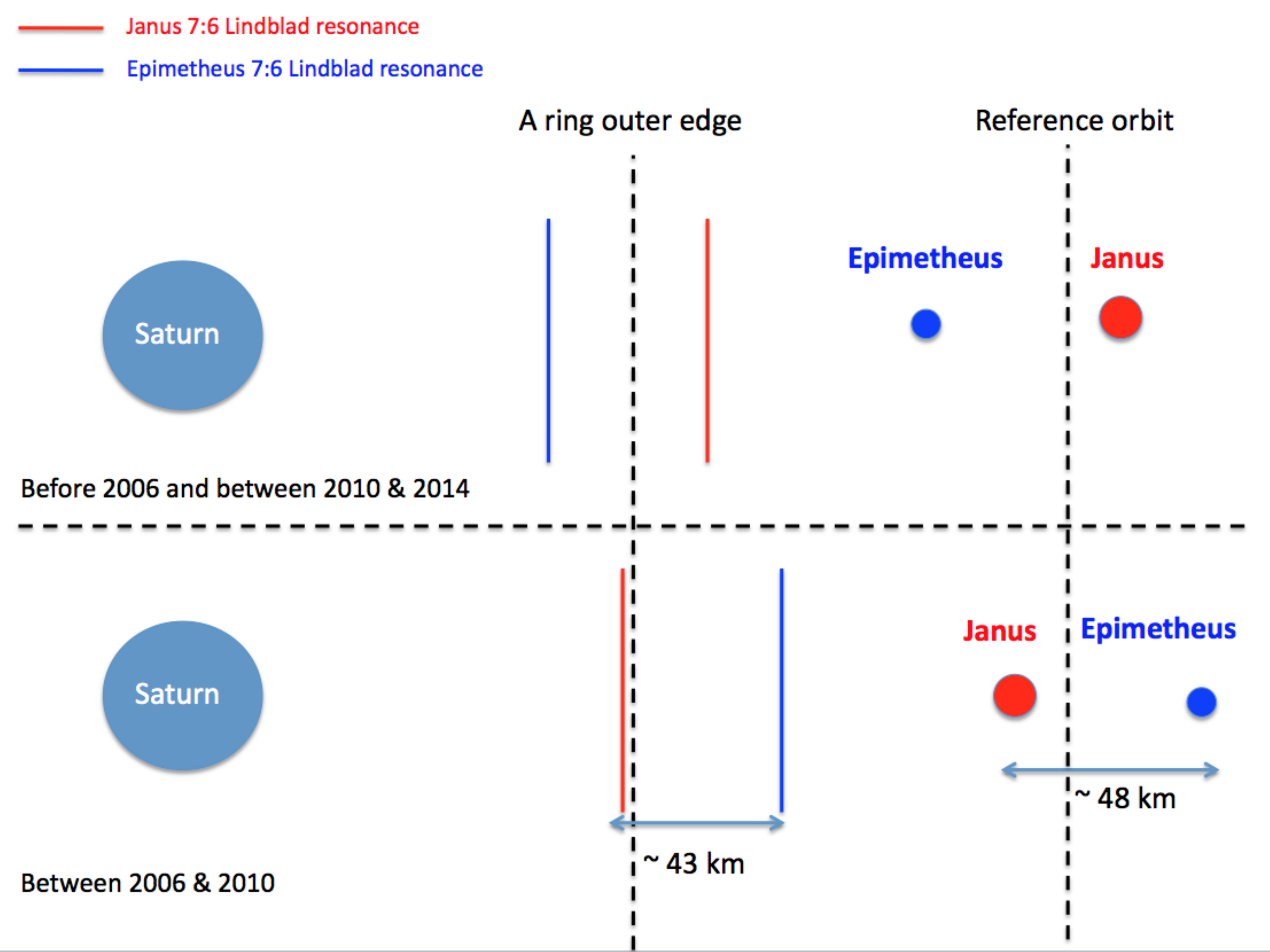}} 
\caption{Cartoon illustrating the effect on the locations of the 7:6 ILRs with Janus and Epimetheus of their periodic orbital exchange. Dashed lines indicate the mean radius of the A ring edge at 136,770~km, and the mean semimajor axis of the satellites. Although the overall system is not shown to scale, the relative orbital positions of the two satellites and the different resonance locations are
drawn to a common scale, as indicated by the scale bars. At each orbital swap, Janus' semimajor axis changes by $\pm21$~km and the corresponding resonance location moves by $\pm19$~km.  The Epimetheus 
resonance likewise follows that of the smaller moon, shifting by $\pm69$~km. It is located $43$~km either exterior or interior to the Janus resonance.}
\label{orbit_swap}
\end{figure}

At the time of Cassini's orbital insertion, Janus was the outer satellite, similar to the situation in 1980/81 when the Voyager flybys occurred. At this time, Janus' 7:6 ILR was located at a radius of 136,785~km, approximately 15~km exterior to the mean radius of the A ring's outer edge. The
Epimetheus 7:6 resonance was located $\sim28$~km interior to the ring edge.  
In January 2006, the first orbital swap occurred and Janus moved to the inner position, with its ILR now located at $136,766$ km, only $\sim4$~km interior to the ring edge.  Four years later, in January 2010, Janus again moved to the outer position and its ILR moved away from the ring edge. Fig.~\ref{orbit_swap} illustrates the changing geometry of the satellites' orbits and the corresponding locations of the two 7:6 Lindblad resonances. 

\cite{spitale09} analyzed a series of 24 Cassini imaging sequences of the outer A ring, obtained between May 2005 and February 2009, in order to characterize the shape of the outer edge.
Starting about 8 months after the co-orbital swap in January 2006, they found a strong $m=7$ radial perturbation rotating at a rate of $518.354 \pm 0.001 \dd$ which closely matched the mean motion of Janus at that time. The mean radial position of the edge was found to be $136,769$ km and the amplitude of the $m=7$ variation was $14.4\pm0.4$ km.
One minimum of this pattern was approximately aligned with Janus, as expected for resonant forcing.
Prior to January 2006, however, the Cassini mosaics showed a more irregular and disorganized appearance, with no clear  periodicity and radial amplitudes as small as $4-5$ km.
\cite{spitale09} ascribed this to a period of readjustment associated with the change in location of the 7:6 resonance in January 2006.
They also noted that a similar readjustment might account for the smaller amplitude ($6.7 \pm 1.5$ km) and slightly lower pattern speed obtained by \cite{porco84} from
Voyager observations, most of which were made only $5$ months prior to the January 1982 coorbital swap.   

This paper is organized as following: In Section 2 we present some theoretical background useful for our data analysis; the data are described in Section 3, our main results are presented in Section 4, and finally, in Section 5 we propose some possible explanations and interpret our results.


\section{\bf Theoretical background}


%
We begin by assuming that the ring edge is perturbed by resonant interactions, specifically by an inner or outer Lindblad resonance (ILR or OLR) and/or normal modes. Following the notation of \cite{nicholson14a, nicholson14b}, for each such resonance or mode the radius of the edge may be described by
\begin{equation}
	r(\lambda,t) = a [ 1-e\cos(m[ \lambda - \Omega_p (t-t_0) - \delta_m ])],     
\label{radius_lambda}
\end{equation}
where $r$ and $\lambda$ are the orbital radius and inertial longitude, respectively, of a ring particle at a given time $t$, $a$ and $e$ are the semi-major axis and orbital eccentricity of streamlines at the ring edge,
$m$ is an integer describing the resonance, $t_0$ is a reference epoch (assumed here to be J2000), and $\Omega_p$ is the appropriate pattern speed. The phase angle $\delta_m$ is the longitude of one of the $m$ minima at time $t = t_0$.  The pattern speed is
given by
\begin{equation}
m\Omega_p= (m-1)n + \dpi, 
\label{patspeed}
\end{equation}
\noindent where $n$ is the ring particle's keplerian mean motion and $\dpi$ is the local apsidal precession rate, as determined by the planet's zonal gravity harmonics. For $m>0$ this equation describes either an ILR or an ILR-type normal mode with $\Omega_p < n$, while for $m<0$ we have either an OLR or an OLR-type mode with $\Omega_p > n$. 

At the outer edge of the A ring, where $a\simeq 136,770$~km, we have $n = 604.22\dd$ and 
$\dpi \simeq 2.95 \dd$, so that the expected pattern speed for a free $m=7$ mode is $\Omega_p = 518.321\dd $. 
This closely matches that of forced perturbations by the Janus 7:6 ILR, which are expected to have
$m=7$ and $\Omega_p \simeq 518.3456\dd$ when Janus is the inner satellite and $\Omega_p=518.2380\dd$ when it is the outer satellite (i.e., after January 2010).
We note that the above frequencies are calculated using the epicyclic theory \citep{renner06}.

To interpret image sequences, it is necessary to shift the central longitude of each image to a common reference epoch $t_0$ using an assumed  pattern speed $\Omega^{*}$. The resulting image is labeled by the comoving longitude coordinate
\begin{equation}
	\theta = \lambda - \Omega^{*}(t - t_0),
\label{theta}
\end{equation}
where $t_0$ is the reference epoch of J2000. 
Usually,  we assemble all the images from a single sequence into a mosaic in radius/comoving longitude ($r,\theta$) coordinates.

Returning to our model in Eq.~(\ref{radius_lambda}), we can rewrite this in terms of comoving longitude by using Eq. (\ref{theta}) to set $\lambda - \Omega_p(t-t_0) = \theta + (\Omega^* - \Omega_p)(t-t_0)$. The predicted radial perturbation becomes
\begin{equation}
	r(\theta,t) = a [ 1-e\cos(m[ \theta + (\Omega^* - \Omega_p)(t-t_0) - \delta_m ])].   
\label{radius_theta}
\end{equation}
If $\Omega^* = \Omega_p$, we have selected the comoving frame to be identical to the frame in which the pattern is stationary, with $m$ lobes and one minimum at $\theta = \delta_m$.
On the other hand, if $\Omega^* \neq \Omega_p$, then individual frames will have their minima offset by $\delta\Omega(t_i-t_0)$, where $\delta\Omega = \Omega_p - \Omega^*$, and the pattern in the mosaic will differ from this simple picture.  
A common approach, and one that we adopt here, is to assume initially that $\Omega^* = n$, the local keplerian mean motion. Assuming also that the observed perturbations are due to a Lindblad resonance located at or near the edge (or to a normal mode), then we have
$\delta\Omega = (-n + \dot{\varpi})/m = -\kappa/m$, where $\kappa$ is the local epicyclic frequency. Over the course of one orbital period ($T=2\pi/n$), the last frame will have its minimum shifted relative to the first frame by $-2\pi\kappa/nm$ or, since $\kappa \simeq n$, by approximately one lobe of the $m$-lobed pattern.
This has the effect of stretching out the pattern so that it appears to have $m-1$ lobes in
$360^\circ$.

A more formal way to see this is to rewrite Eq. (\ref{radius_theta}) to eliminate the explicit time dependance, by noting that
\begin{equation}
	t-t_0 = (\lambda - \theta)/\Omega^*,     
\label{time_dep}
\end{equation}
and
\begin{equation}
	\delta\Omega = \Omega_p-\Omega^* = [(m-1)n - m\Omega^* + \dpi]/m. 
\label{}
\end{equation}
If we now set $\Omega^* = n$, this expression simplifies to
\begin{equation}
	\delta\Omega = \frac{-n+\dot{\varpi}}{m} = -\frac{\kappa}{m},   
\label{}
\end{equation}
where again $\kappa$ is the local epicyclic frequency, so that we have 
\begin{equation}
	\delta\Omega(t-t_0) = -\frac{\kappa}{mn}(\lambda - \theta).   
\label{delta_omega}
\end{equation}
Substituting this into Eq.~(\ref{radius_theta}), we have
\begin{equation}
	r(\theta,\lambda) = 
	a \left(1 - e \cos \left[(m-\frac{\kappa}{n}) \theta + \frac{\kappa}{n}\lambda - m\delta_m\right]\right).  
\label{radius3}
\end{equation}

Since $\kappa/n \simeq 1+ O(J_2[R_p/a]^2)$, where $J_2$ is the second zonal gravity harmonic and $R_p$ is Saturn's equatorial radius,  the pattern of radial perturbations for $m>0$ ({\it i.e.,} for ILR-type modes) will have approximately $m-1$ lobes in comoving coordinates with one minimum at $\theta_{min} \simeq (m\delta_m - \lambda)/(m-1)$. 
However for $m<0$ ({\it i.e.,}  OLR-type modes), the pattern has approximately $|m| +1$ lobes, with one minimum at $\theta_{min} \simeq (|m|\delta_m - \lambda)(|m|+1)$.  Note that it is, in general, impossible to determine from a single image sequence that exhibits $m'$ radial
lobes whether this perturbation is due to an ILR with $|m| = m'+1$ or an OLR with $|m|= m'-1$.
Only by combining mosaics obtained at different times can we determine $\Omega_p$ and thus establish the true identity of the resonance or normal mode.

For the predicted $m=7$ perturbation associated with the Janus 7:6 ILR, we thus expect to
see a 6-lobed radial pattern in mosaics assembled with $\Omega^* = n$, with a phase that
depends on the inertial longitude of the particular image sequence, as well as the orbital  longitude of Janus.

Returning to Eq.~(\ref{radius_theta}), for an arbitrary value of $\Omega^*$ we have
\begin{equation}
	r(\theta,\lambda) = a \left( 1-e\cos \left[m\theta + \frac{m\Omega^* - (m-1)n-\dpi}{\Omega^{*}}(\lambda - \theta) - m\delta_m\right]\right),    
\label{}
\end{equation}
which can be simplified to
\begin{equation}
	r(\theta,\lambda) = a \left( 1-e\cos\left[\frac{m\Omega_p}{\Omega^{*}}\theta +         \frac{m(\Omega^{*} - \Omega_p)}{\Omega^*}\lambda - m\delta_m\right]\right).
\label{}
\end{equation}
So, if the inertial longitude of the images,\ $\lambda$ is constant, then the azimuthal wavelength of the radial pattern  is given by
\begin{equation}
\Lambda = \frac{2 \pi \Omega^{*}}{m\Omega_p} = \left\{ 
\begin{array}{l l}
  2\pi/m & \quad \text{if $ \Omega^{*} = \Omega_p$}\\
  2\pi n/[(m-1)n + \dpi] \simeq 2\pi/(m-1) & \quad \text{if $\Omega^{*} = n$}.\\ 
\end{array}  \right .
\label{strem_m_1}
\end{equation}
Note that for $\Omega^* = \Omega_p$ and $\Omega^* = n$ we recover the special cases discussed above.  This expression enables us to reinterpret periodic radial perturbations observed in individual movies, even if the assumed pattern speed $\Omega^*$ turns out
to differ from the true pattern speed $\Omega_p$.
Note that if the pattern speed is not given by Eq. (\ref{patspeed}), there need not necessarily be an integer number of wavelengths in $360$ degrees.
%

Finally, we consider briefly the question of whether, when the resonance moves outwards by $19$ km, the mean radius of the ring edge 
should do likewise.
A simple calculation of the time taken for an unconstrained ring to spread a distance $\Delta_r$ via collisional interactions yields the expression:
\begin{equation}
	t = \frac{\Delta_r^{2} }{\nu},  
\label{strem_m_2}
\end{equation}
where $\nu$ is the effective kinematic viscosity.
A rough estimate of $\nu$ is provided by the standard expression for non-local viscosity \citep{gold82}
\begin{equation}
	\nu = \frac{2c^{2} \tau }{n(1+\tau^{2})},  
\label{strem_m_2}
\end{equation}
where $c$ is the RMS velocity dispersion.
Adopting a plausible value of $c=0.1$ cm/s \citep{tiscareno07}, corresponding to a ring thickness $H\sim c/n \simeq 8$ m, and $\tau \sim 0.5$, 
we find that $\nu \simeq 65 ~cm^2s^{-1}$ and $t=1800$ years for $\Delta_r = 19$ km.
We therefore expect negligible radial spreading ($\Delta_r \sim 1$ km) in 4 years.

Our observations confirm this.
The most accurate absolute radii come from the occultation data, and the models in Table 8 indicate that the mean radius of the A ring's
outer edge changed by only $2.2 \pm 1.1$ km between 2006-2009 and 2010-2013.
Given the possibility of unregonized systematic errors in the analysis (e.g., missing terms in our model of the edge), we do not consider this shift to be significant.


\section{\bf Data analysis}


\subsection{Occultation data}

In previous studies of the B ring edge and of sharp-edged features in the C ring  \citep{nicholson14a,nicholson14b} we have
assembled complete sets of radial optical depth profiles for the rings from both radio and stellar occultation data.  This combined dataset contains over 170 cuts across the outer edge of the A ring, of which 15 were obtained prior to the orbital swap in January 2006, 118 between January 2006 and the succeeding swap in January 2010, and 35 between January 2010 and January 2014. (A small
number of occultations acquired since January 2014 have not yet been fully reduced and are not included here.) The radial resolution of these profiles varies from a few 10s of meters to
$\sim1$~km, and the radii are computed to an internal accuracy of $\sim200$~m. Any systematic errors in the overall radius scale are believed to be less than  300~m \citep{nicholson14a}.

We determine the radius of the outer edge of the A ring in each profile by fitting a
standard logistic profile to the measured transmission as a function of radius to obtain the radius at half-light.  Since the A ring edge is quite sharp, the uncertainty in the estimated
radius is comparable to the resolution of the dataset, or $<1$~km. The corresponding event time, as recorded on the spacecraft or at the ground station for radio occultations, is back-dated to account for light-travel time from the rings to yield the observation time, $t$.
Together with the calculated inertial longitude of each cut, $\lambda$, derived from the
spacecraft ephemeris and the stellar or Earth direction with respect to Cassini, this completes our data set.

In order to determine the parameters of the edge, we fit the measured radii with Eq.~(\ref{radius_lambda}), using a nonlinear least-squares routine based on the Leavenburg-Marquardt algorithm.  This well-tested routine is the same as that used by \cite{nicholson14a} to fit the B ring edge.  Our nominal model includes an $m=7$ ILR-type perturbation, due to the Janus 7:6 resonance, plus one or more normal modes as required. The latter are first identified with 
the aid of a spectral-scanning routine, where, for an assumed value of $m$, multiple fits are carried out for a range of pattern speeds in the neighborhood of that predicted by 
Eq.~(\ref{patspeed}). In order to minimize aliasing, we first remove any already-identified
modes from the data before scanning for additional modes, but in our final fits we solve for
the parameters of all identified modes simultaneously.
For more details and examples, as well as a complete list of observations, the interested reader is referred to \cite{nicholson14a}.


\subsection{Imaging data}

The imaging data were obtained by the ISS Narrow-Angle Camera (NAC) on board the Cassini spacecraft between 2006 and 2014 (see Table \ref{tab_movie_iss}).
This table lists the 37 sequences analyzed here, including the identification (ID), which is the 4 first digits of the initial image in the sequence, the date, the time range in hours and the inertial longitude range in degrees. 

For the majority of our
imaging data, referred to in Cassini catalogs as ``movies'', each image in a sequence is targeted at approximately the same inertial longitude, $\lambda$ with continuous coverage over a period of time which is comparable to the local orbital period $T$. For the A ring edge, $T = 14.3$~hours, while for the
F ring (the target of many of our  movies) it is 14.8~hours. In some cases, a single movie is composed of two or more shorter observations spread over a period of several days. 

\begin{figure}[!h]
\centerline{\includegraphics[angle=0,totalheight=5cm,trim= 0 0 0 0]{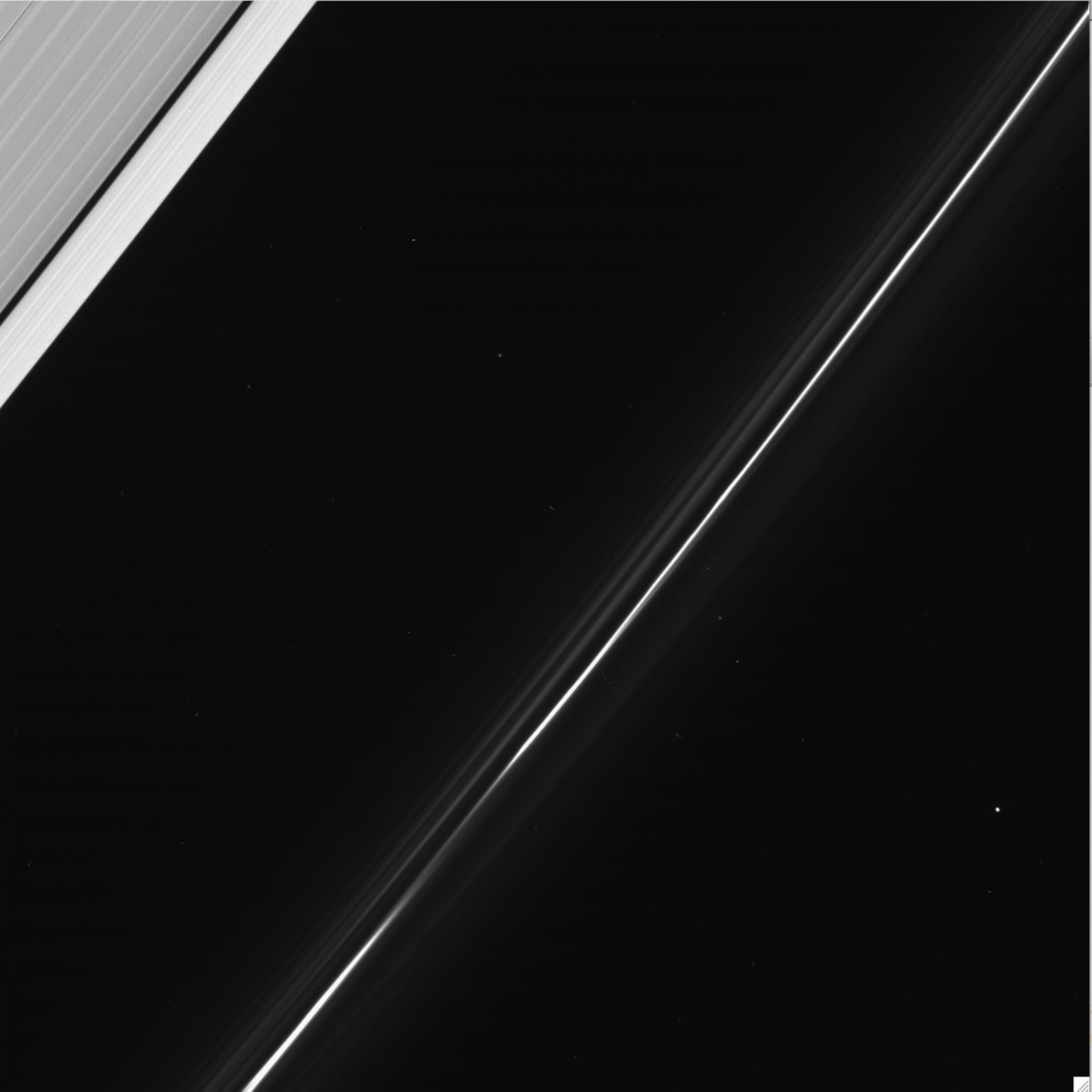}} 
\caption{%
Cassini ISS image N1616515163 taken in March 2009 by the ISS instrument on board the Cassini spacecraft, as part of movie number 23. 
}%
\label{Cassini_image_1615}
\end{figure}

As an example, we choose movie number 23, which has 208 separate images taken over 12.88 hours and spans a $45^{\circ}$ range of inertial longitude. One of these images is shown in Fig. \ref{Cassini_image_1615}.
Here, as in many F movies (sets of images of the F ring), we can observe a segment of the outer edge of the A ring. 

We first navigated each image using background stars,
using the technique described in detail by \cite{murray05}.

We then measured the radii of the outer edge of the A ring. 
First, the images were reprojected onto a longitude-radius grid, and then the pixel values at each longitude (i.e., for each vertical line in the reprojected image) were differentiated and fit to a gaussian to determine the radial location of the maximum slope in the observed brightness at that longitude. This method gives a robust sub-pixel determination of the edge location; radii determined at neighboring longitudes are typically consistent to within 0.1 pixels. 
For each reprojected image, we have up to $1000$ radius measurements, and for each movie, around $10^5$ data points versus inertial longitude and time.

\begin{figure}[!h]
	\center\includegraphics[angle=270,width=0.65\columnwidth]{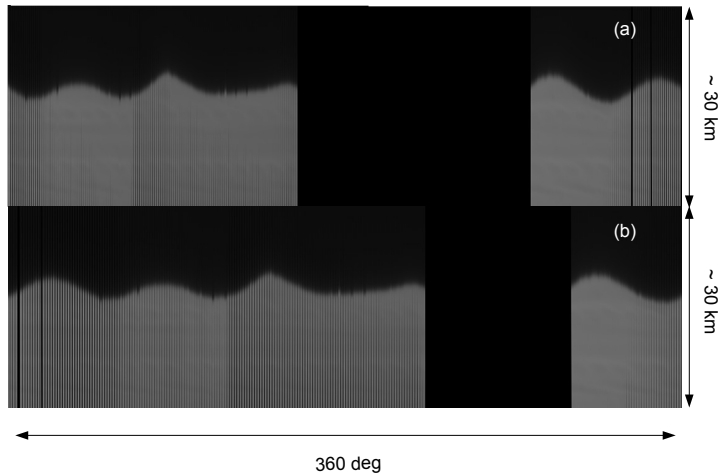}
\caption{%
Mosaics of the A ring edge assembled from a set of about 200 ISS Cassini images from movie number 23 (see Table \ref{tab_movie_iss}) taken in March 2009. Panels (a) and (b) are made in the Janus frame ($\Omega^{*} = n_J$) and the local keplerian frame ($\Omega^{*} = n$), respectively.
Note that the gap in coverage is smaller in the local frame. Time increases from right to left, while co-moving longitude $\theta$ increases from left to right.
}%
\label{mos_before}
\end{figure}

\begin{figure}[!h]
	\center\includegraphics[angle=270,width=0.65\columnwidth]{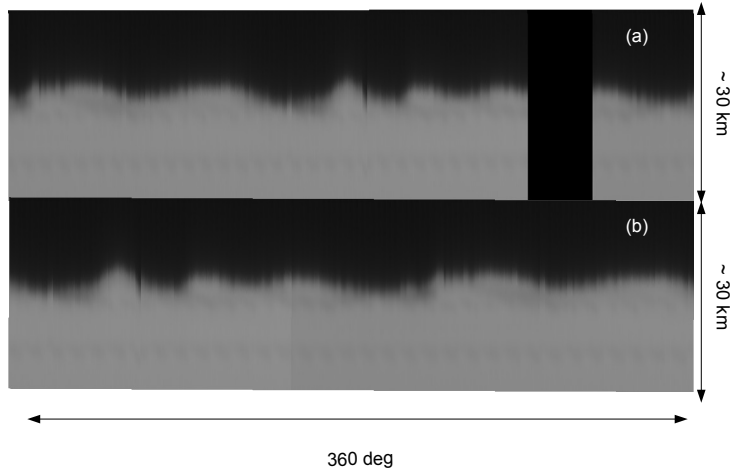}
\caption{%
Mosaics of the A ring edge assembled from a set of about 200 ISS Cassini images from movie number 33 (see Table \ref{tab_movie_iss}) taken in August 2013. Panels (a) and (b) are made in the Janus frame ($\Omega^{*} = n_J$) and the local keplerian frame ($\Omega^{*} = n$), respectively. 
There is no gap in the local frame, in this case, as the movie's duration exceeds one orbital period.
}%
\label{mos_after}
\end{figure}

Finally, we assembled all reprojected images for a given movie sequence into two mosaics in radius-longitude space, using $\Omega^{*} = n$ and $\Omega^{*} = n_J$. 
Figures \ref{mos_before} and \ref{mos_after} show examples of these mosaics, from before and after 2010.


%
%


%
%

We analyze these data in two different ways : 

\paragraph{Famous software: } 

As a first step, we plot the data for each movie in two different frames: in the Janus frame we plot  radius versus Janus' corotating longitude ($\lambda_I - \lambda_J$), while in the local or keplerian frame we plot radius versus local corotating longitude $\theta$ with $\Omega^{*} = n$, following Eq.  \ref{theta}.
Then, after decimating (pick one sample out of ten) the raw data from $10^5$ samples to $10^4$ samples per movie, we carry out a frequency analysis of the edge using the Famous software (ftp://ftp.obs-nice.fr/pub/mignard/Famous/).
The latter uses a non-linear least-squares model to fit the data with a series of sinusoidal signals of arbitrary amplitude, frequency and phase.
For a multi-periodic signal, once a frequency has been identified, the program removes this component from the signal; 
this process is then repeated a specified number of times.
This method requires us first to choose a certain pattern speed, 
but the frequencies are not constrained \textit{a priori} (i.e., $\Lambda$ need not be $2\pi/m$ for integer $m$). 
In each fit, data from only a single movie are analyzed. 

\paragraph{Root Mean Square (RMS): }
Once candidate values of $m$ are determined by the Famous fits, we next attempt to extend these models over many movies (up to 4 years) by determining the phase of the periodic signal $\Phi$ in each movie (see Eq. \ref{phi_rms} below).
In this case, we assume a particular value of $m$ and look for the best-fitting value of $\Omega_p$ which minimizes the RMS phase difference, $\Psi$, between movies. 
Our procedure is as follows:
first, for a range of pattern speeds, we calculate the corotating longitude of each data point for every trial value of the pattern speed as $\theta_i = \lambda_{i} - \Omega_p(t_i - t_0)$, where $t_0$ is the J2000 epoch, chosen as a reference time.
Then for each movie we compute the sum of cosine and sine terms as

\begin{equation}
	C = \frac{1}{N} \sum\limits_{i=1}^N {\Delta r_i \cos(m\theta_i)},  ~~~~~~~~  S = \frac{1}{N} \sum\limits_{i=1}^N {\Delta r_i \sin(m\theta_i)},
\label{cs_rms}
\end{equation}
where $\Delta_r$ is the difference in radius from the mean value of the edge (taken as $136,770.0$ km) and $N$ is the number of data points in the movie (typically $\sim~10^4$). 
Then we have for every value of $\Omega_p$, estimates of the phase $\Phi$ and amplitude $A$ from:

\begin{equation}
	\Phi  = \arctan(-S/C) 
\label{phi_rms}
\end{equation}
and
\begin{equation}
	A = 2\sqrt{C^2 + S^2}.
\label{amp_rms}
\end{equation}

\noindent This is equivalent to fitting the measured radii with a model  :

\begin{equation}
	\Delta_r = \bar{\Delta}_r + A\cos(m\theta + \Phi) + \text{Noise}.
\label{Delta_r_rms}
\end{equation}

\noindent The classical mean square phase variation is given by : 

\begin{equation}
	\sigma^{2} = \frac{1}{M-1} \sum\limits_{j=1}^M ({\Phi_j - \bar{\Phi})^2},
\label{rms}
\end{equation}
where $M$ is the number of movies and $\bar{\Phi}$ is the mean phase over all the movies for a given value of the pattern speed.

\noindent However, this formula creates problems because of the periodicity of the phase, which can lead to ambiguities of $\pm2\pi$ in the phase difference.  
To avoid this problem, we instead calculate a modified form of the mean square variation given by: 

\begin{equation}
	\Psi^2 = \frac{2}{M(M-1)} \sum\limits_{i=1}^{M-1}  \sum\limits_{j=0}^{i-1}  \sin^2(({\Phi_i - \Phi_j)/2)},
\label{rms_psi}
\end{equation}
where the sum is over all distinct pairs of movies.
This measure of the RMS variation treats phase differences of $2\pi$ as equivalent to $0$.
Examples of this method are given in the next section.

\begin{table}[H]
\vskip4mm
\centering
\begin{tabular}{|*{7}{c|}}
\hline
  Movie & ID & Date & Time range (hours)  &   Inertial longitude $\lambda_I$  (deg)  \\
\hline
  1 &  1538 & 2006-271 & 36.50 &  78.54 - 274.00   \\
\hline
  2 & 1539 &  2006-289 &  7.75 &  261.99 - 271.24  \\
\hline
  3 &  1541 &  2006-304 &  29.77 & 91.52 - 102.93   \\
\hline  
  4 & 1542 &  2006-316 & 30.50  & 91.23 - 107.98   \\
\hline
  5 & 1543 & 2006-329 &  13.94 & 290.03 - 301.36   \\
\hline
  6 &  1545 & 2006-357 & 15.48  &  282.05 - 294.51 \\
\hline
  7 & 1546 & 2007-005 &  13.26 &  247.47 - 260.89  \\
\hline
  8 & 1549 & 2007-041 &  30.71 & 239.37 - 256.00   \\
\hline
  9 & 1551 & 2007-058 &  15.44 &  193.98 - 210.71 \\
\hline
  10 & 1552 & 2007-076 &  16.80 &  201.69 - 218.25  \\
\hline
  11 &  1554 & 2007-090 &  12.54 &  171.49 - 182.96 \\
\hline
  12 & 1557 & 2007-125 &  14.05 &  172.44 - 187.03  \\
\hline
  13 &  1577 & 2007-365 &  13.48 &  163.73 - 176.44  \\
\hline
  14 &  1579 & 2008-023 &  13.06 & 138.29 - 150.30   \\
\hline
  15 & 1598 & 2008-243 &  12.89 & 106.25 - 120.58   \\
\hline
  16 &  1600 & 2008-260 &  12.60 & 126.53 - 310.34  \\
\hline
  17 &  1601 & 2008-274 & 34.44  &  282.69 - 300.04  \\
\hline
  18 & 1602 & 2008-288 &  11.94 &  280.30 - 286.57 \\
\hline
  19 &  1604 & 2008-303 &  10.63 & 197.52 - 303.86   \\
\hline
  20 & 1610  & 2009-011 &  11.19 &  123.01 - 136.01 \\
\hline
  21 & 1612  & 2009-041 &  8.39 & 165.55 - 194.61   \\
\hline
  22 &  1615 & 2009-070 &  13.41 &  146.00 - 194.07  \\
\hline
  23 &  1616  & 2009-082 & 12.88  & 145.29 - 191.96  \\
\hline
  24 & 1618  & 2009-106 &  9.86 & 236.14 - 250.23  \\
\hline
  25 & 1620  & 2009-130 & 10.79  & 248.91 - 266.47  \\
\hline
  26 & 1627 & 2009-211 & 12.66  &  231.27 - 241.75  \\
\hline
  27 &  1729 & 2012-289  & 16.01  & 107.03 - 338.05   \\
\hline
  28 &  1743 & 2013-077 &  11.18 &  4.82 - 11.79  \\
\hline
  29 &  1746 & 2013-126 &  14.43 & 237.92 - 244.13   \\
\hline
  30 &  1748 & 2013-147 & 14.02  & 263.26 - 267.28   \\
\hline
  31 &  1750 & 2013-171 &  15.81 & 42.06 - 241.38   \\
\hline
  32 & 1755  & 2013-232 &  14.83 &  237.09 - 243.77  \\
\hline
  33 & 1756  & 2013-236 &  14.83 &  274.61 - 280.57  \\
\hline
  34 &  1757 & 2013-250 &  7.86 &  252.88 - 257.71  \\
\hline
  35 &  1760 & 2013-291 &  15.81 &  274.48 - 113.63  \\
\hline
  36 &  1776 &   2014-103  & 15.84 &  27.86 - 208.77  \\
\hline
  37 &  1782 & 2014-173 & 13.24  &  66.42 - 75.53  \\
\hline
\end{tabular}\\
\caption{List of all ISS movies analyzed between 2006 and 2014.}
\label{tab_movie_iss}
\end{table}

\newpage


\section{Results}


\subsection{Analysis of the data taken between 2006 and 2010}

\subsubsection{Occultation data}
 
\begin{figure}[!h]
\centerline{\includegraphics[angle=0,totalheight=15cm,trim= 0 0 0 0]{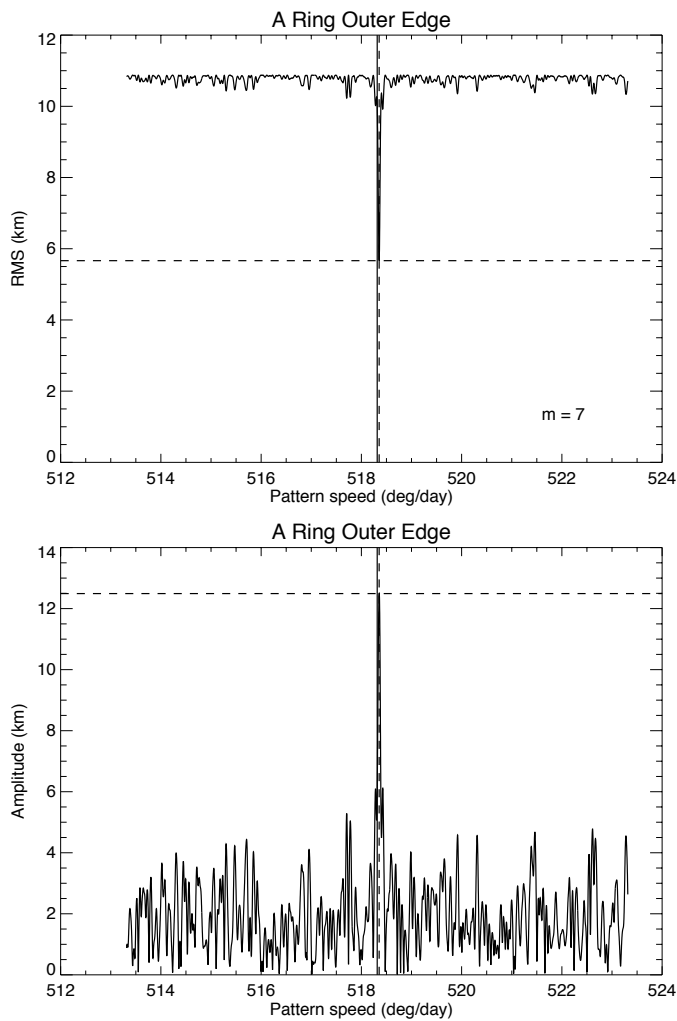}} 
\caption
{Identification of the $m=7$ radial perturbation on the outer  edge of the A ring in occultation data obtained between January 2006 and July 2009. The upper panel
shows the RMS residuals with respect to Eq.~(\ref{radius_lambda}), as the pattern speed $\Omega_p$ is scanned across the expected rate of $518.321\dd$,
while the lower panel shows the corresponding radial amplitude $ae$.  The vertical dashed line indicates the rate that provides the best fit to the data (in this case
$\Omega_p = 518.354\dd$), while the solid line indicates the predicted rate for an $m=7$ ILR that falls at the fitted semimajor axis of the edge. Horizontal
dashed lines indicate the minimum RMS residual of 5.6~km and corresponding maximum amplitude of 12.5~km.}
\label{occdata_m7_before}
\end{figure}

The period between January 2006 and July 2009 contains the bulk of our occultation datasets, with
a total of 118 useable  profiles. The RMS scatter in the measured radii relative to a circular ring edge is
10.7~km, which is comparable  to the amplitude of the variations reported by \cite{spitale09} in imaging data from this period. 
As expected, the data show a strong $m=7$ signature, with a radial amplitude
of $\sim12.5$~km and a pattern speed very close to that of Janus' mean motion at the time, as illustrated in 
Fig.~\ref{occdata_m7_before}.  Our adopted best fit, as documented in Table 8, yields a pattern speed $\Omega_p = 518.3544\pm0.0008\dd$ and an amplitude of $12.9\pm0.3$~km.
In this time interval, $n_J=518.3456\dd$.

\begin{figure}[!h]
\centerline{\includegraphics[angle=0,totalheight=15cm,trim= 0 0 0 0]{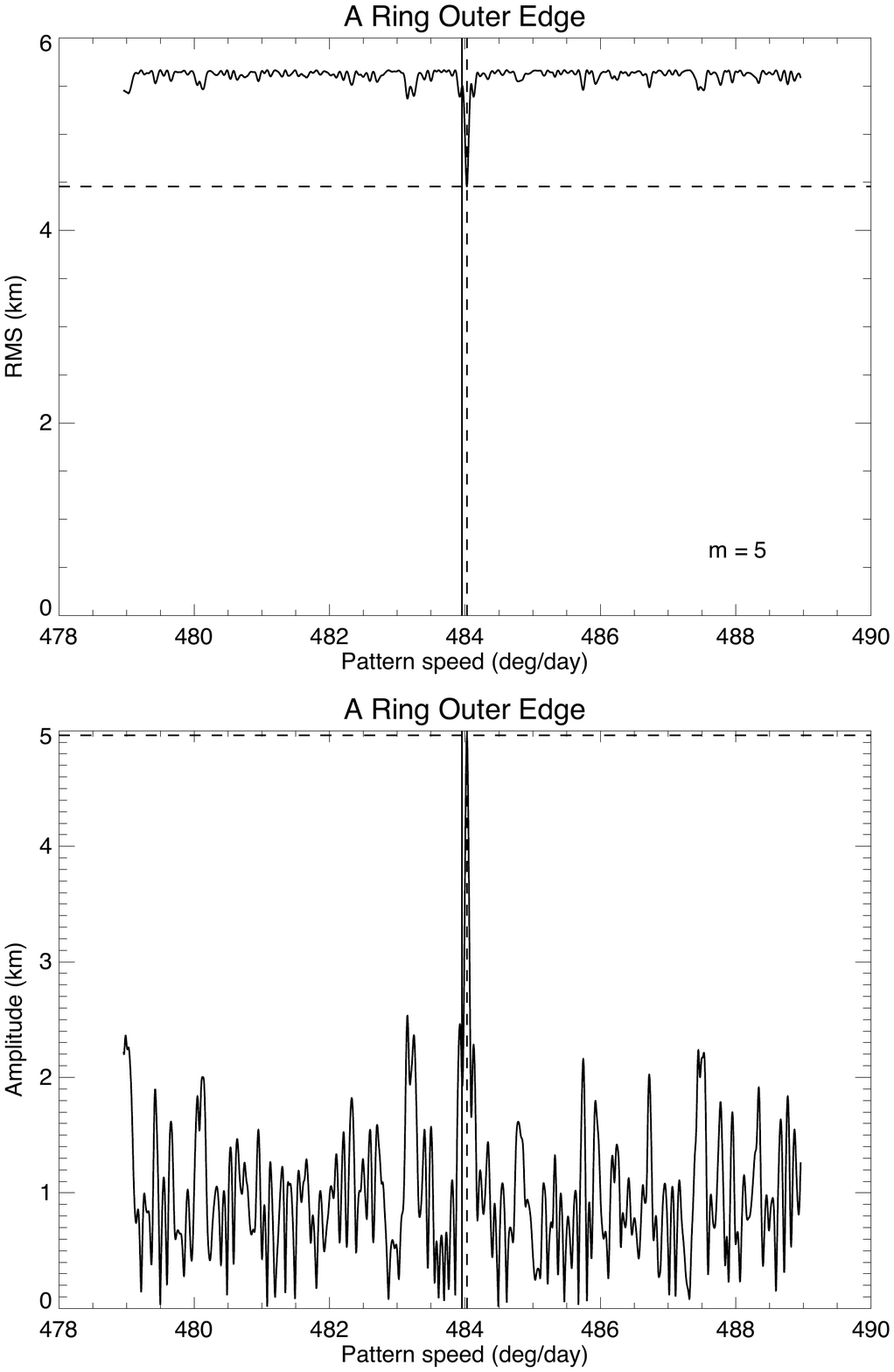}} 
\caption
{Identification of an $m=5$ normal mode on the outer  edge of the A ring in occultation data obtained between January 2006 and July 2009. The format is the same
as that of Fig.~\ref{occdata_m7_before}. The vertical dashed line indicates the rate that provides the best fit to the data (in this case
$\Omega_p = 484.027\dd$), while the solid line indicates the predicted rate for an $m=5$ ILR that falls at the fitted semimajor axis of the edge, or $483.957\dd$.
The maximum amplitude is 5.0~km. For these fits the best-fitting $m=7$ model from  Fig.~\ref{occdata_m7_before} was first subtracted from the raw data.}
\label{occdata_m5_before}
\end{figure}

\begin{figure}[!h]
\centerline{\includegraphics[angle=0,totalheight=10cm,trim= 0 0 0 0]{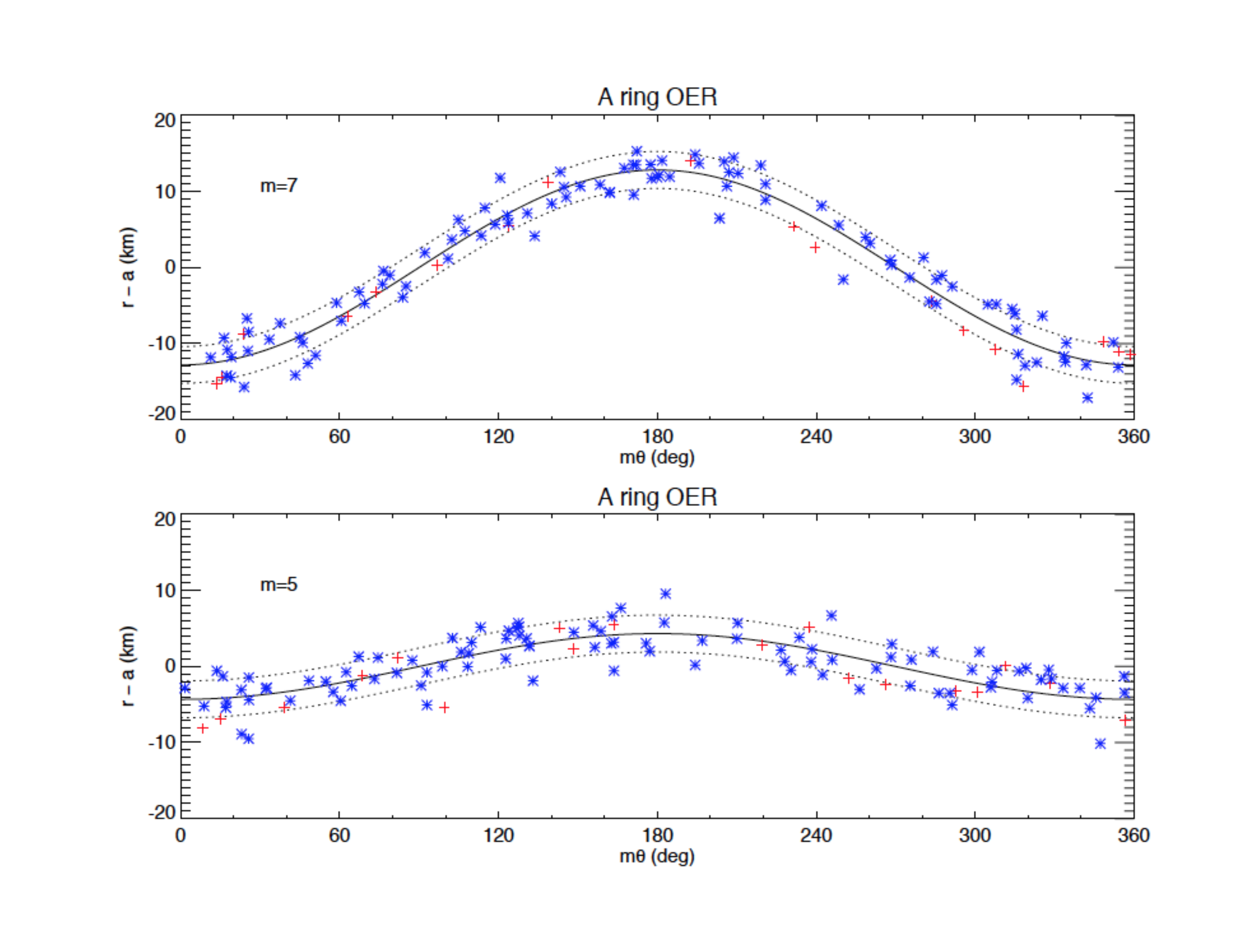}} 
\caption
{Measured radii of the outer  edge of the A ring in occultation data obtained between January 2006 and July 2009, relative to the mean semi-major axis $a = 136,770.5$~km.
The upper panel shows the radii plotted vs the argument  $m\theta = 7[\lambda - \Omega_p (t-t_0) - \delta_7]$, after subtraction of the best-fitting models for all modes other than $m=7$. The pattern speed $\Omega_p = 518.354\dd$.  The lower panel shows the same data, but plotted vs the argument  $5[\lambda - \Omega_p (t-t_0) - \delta_5]$,
again after subtraction of all other modes. In this case, $\Omega_p = 484.021\dd$.
The solid curves show the best-fitting $m=7$ and $m=5$ models from  Table 8, with dotted lines indicating the
RMS residuals with respect to the overall model.}
\label{occresids}
\end{figure}

But the post-fit RMS residual with a pure $m=7$ model is 5.6~km, which while much less than that of the raw data is still large compared to our measurement errors of $\sim 300$~m.  Further experiments with our spectral-scanning program unexpectedly revealed the existence of a substantial $m=5$ ILR-type mode, with an amplitude of $\sim5$~km (see Fig.~\ref{occdata_m5_before}). Least-squares fits confirm the reality of this mode, and our adopted fit yields a corresponding pattern speed of $484.021\pm0.004\dd$, and amplitude $5.1\pm0.6$~km. 
When both modes are included, the post-fit residuals are reduced to 4.4~km.  Further investigation indicated the presence of additional, weaker ILR-type normal 
modes with $m=3$, 4, 6, 8, 9, 10 and 18 and amplitudes of 1.5--3.1~km (not plotted here). 
Our final adopted fit includes all nine modes, as documented in Table 8, and has an RMS residual of 1.8~km.
In Fig.~\ref{occresids}, we plot the individual measured radii vs our best-fitting $m=7$ and $m=5$ models. 
Based on preliminary results from our analysis of the imaging data (see below), we also searched for evidence of OLR-type modes, but did not find anything significant.

Several lines of evidence suggest that the seven weaker modes are real, despite their small amplitudes.
First, we observe each of these modes to move at the pattern speed expected for such normal modes, i.e., at the same rate as if they were forced by a Lindblad resonance, as given by Eq. (\ref{patspeed}) above.
Second, in each case the fitted pattern speed
corresponds to a resonant radius $\ares$ that is located only a few km \textit{interior} to the mean ring edge, as expected for the resonant-cavity modes  \citep{nicholson14a}.
Third, we note that, if these modes simply represented accidental minima in the radius residuals, then one would expect similar accidental OLR-type modes,
which we do not see.

\subsubsection{Imaging data}

\begin{figure}[!h]
	\center\includegraphics[angle=0,width=0.75\columnwidth]{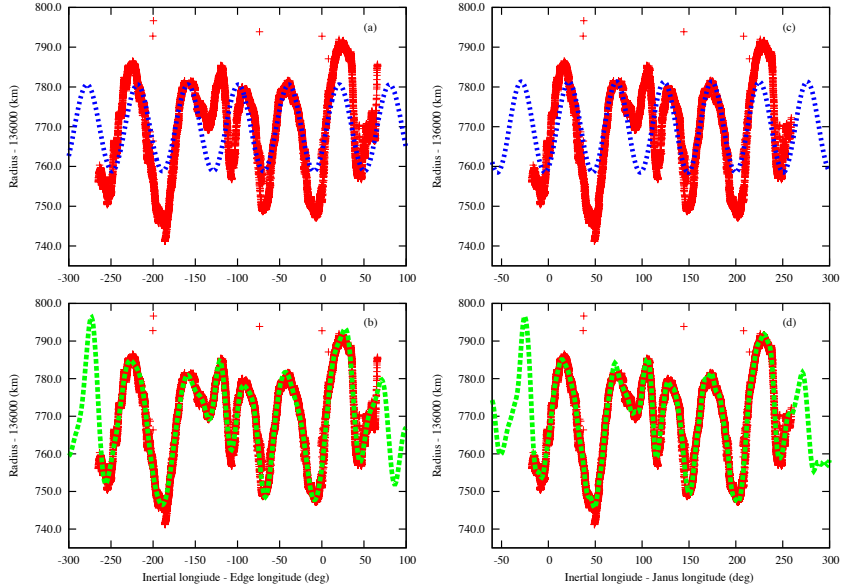}
\caption{%
Red curves are data from Cassini images for movie number 7 (see Table \ref{tab_movie_iss}) obtained in January 2007,
mosaicked in the keplerian frame (left panels) and in the Janus frame (right panels).
Panels (a) and (c) show the dominant frequency in each frame (blue curves), while panels (b) and (d) show the combined signal of the all fitted components (green curves). Note that points are the data and the curves are the least-squares models.
}%
\label{famous_before}
\end{figure}

When the imaging mosaics for the 2006-2009 movies are assembled using $\Omega^* = n_J= 518.34  \dd$, we see in almost all cases an $m=7$ pattern.
This pattern is as expected, due to the 7:6 ILR with Janus, and as described above by Eq. (\ref{radius_theta}).
If instead the data from an F-movie (which have nearly constant $\lambda$) are mosaicked with $\Omega^{*} = n = 604.22\dd$, then typically we see a 6-lobed pattern whose phase depends on $\lambda$ as well as $\delta$, consistent with Eq. (\ref{radius3}) above. 
This behavior is illustrated for one image sequence in Fig. \ref{famous_before}, with the corresponding least-squares fit parameters given in Tables \ref{tab_famous_edge_before} and \ref{tab_famous_janus_before}.
This figure shows the radius measurements, along with the best-fitting sinusoidal models. Panels (a) and (b) show the measured radii for a movie obtained in January 2007, 
assembled using $\Omega^{*} = n$, the local keplerian mean motion. 
Panel (a) shows the dominant frequency only, while panel (b) shows the sum of all 9 fitted components to the edge.
We note that the dominant frequency has approximatively 6 lobes over 360 degrees. The other components, however, are not negligible.  
Panels (c) and (d) show data from the same movie sequence, but processed using $\Omega^{*} = n_J$. 
In this case the dominant frequency is a sinusoid with 7 lobes over 360 degrees.

\begin{table}[H]
\vskip4mm
\centering
\begin{tabular}{|*{5}{c|}}
\hline
  Component & Amplitude (km) & Period (deg) & Phase (deg)  &  $m$  \\
\hline
  1 & 10.89 & 59.79  & 235.93 & 6.02   \\
\hline
  2 & 6.3 & 88.64 & 205.13 &  4.06  \\
\hline
  3 & 5.9 & 47.95 & 244.38  &  7.5 \\
\hline
\end{tabular}\\
\caption{Frequency analysis for the data in Fig. \ref{famous_before}, with $\Omega^{*} = n$.}
\label{tab_famous_edge_before}
\end{table}
\begin{table}[H]
\vskip4mm
\centering
\begin{tabular}{|*{5}{c|}}
\hline
  Component & Amplitude (km) & Period (deg) & Phase (deg)  &  $m$  \\
\hline
  1 & 11.41 & 50.84  & 205.25 &  7.08 \\
\hline
  2 & 6.38 & 75.64 & 294.18 & 4.75   \\
\hline
  3 & 5.66 & 41.30 & 219.75  &  8.7  \\
\hline
\end{tabular}\\
\caption{Frequency analysis for the data in Fig. \ref{famous_before}, with $\Omega^{*} = n_J$.}
\label{tab_famous_janus_before}
\end{table}

Tables \ref{tab_famous_edge_before} and \ref{tab_famous_janus_before} show the parameters of the largest three fitted components seen in Fig. \ref{famous_before}, with the frequency expressed in terms of the equivalent value of $m$\footnote{Note that the least-squares fitting program does not assume that the wavelength is a submultiple of 360\textdegree, corresponding to an integer value of $m$}.

\begin{figure}[!h]
	\center\includegraphics[angle=0,width=0.6\columnwidth]{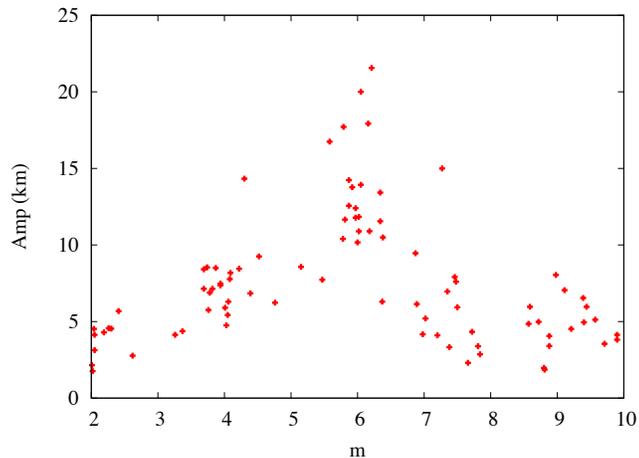}
\centerline{} 
\caption{%
Results of frequency analysis using $\Omega^{*} = n$ for all the movies obtained between 2006 and 2009.
We see the dominance of modes with $m=6$ and $m=4$.
}%
\label{m_amp_before_edge}
\end{figure}

\begin{figure}[!h]
	\center\includegraphics[angle=0,width=0.6\columnwidth]{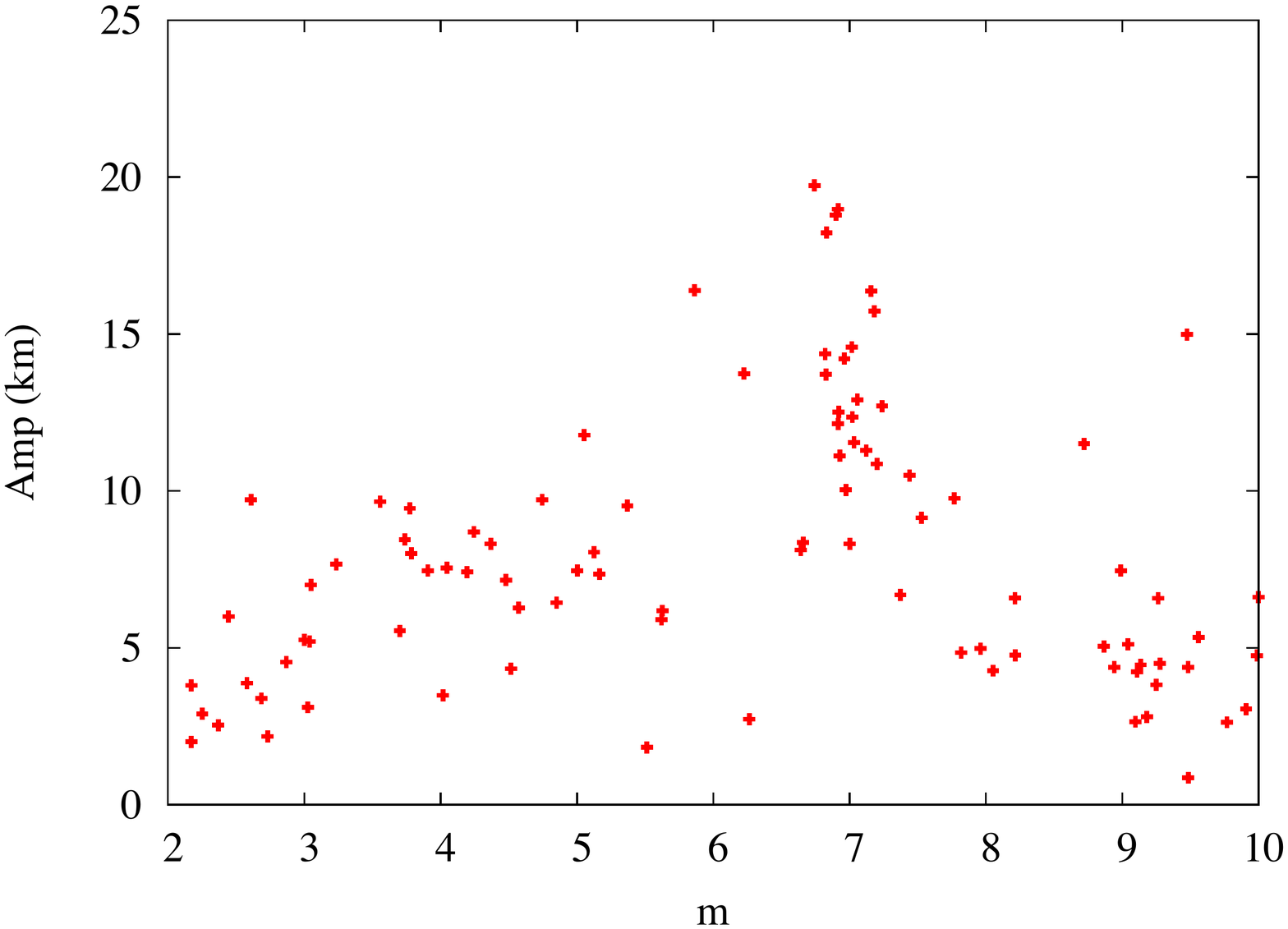}
\centerline{} 
\caption{%
Results of frequency analysis using $\Omega^{*} = n_J$ for all the movies obtained between 2006 and 2009.
We see the dominance of modes with $m\simeq7$, as expected.
}%
\label{m_amp_before_janus}
\end{figure}

We note here that Equation (\ref{strem_m_1}) predicts that a signature with wavelength $\Lambda$ in the local frame should appear with wavelength $(n_J/n)\Lambda \sim  (6/7)\Lambda$ in the Janus frame, which agrees with what we see in Tables \ref{tab_famous_edge_before} and \ref{tab_famous_janus_before}.
We summarize our least-squares results for all of the image sequences obtained in 2006-2009 in Fig. \ref{m_amp_before_edge} and Fig. \ref{m_amp_before_janus}, in the form of scatter plots of amplitude versus frequency 
(expressed as the equivalent value of $m$).
A complete list of fit parameters is given in Table \ref{table_famous_edge_before} in the Appendix. 
Two sets of results are shown here, for $\Omega^{*} = n$ (the local frame) and $\Omega^{*} = n_J$.
The amplitude of the dominant $m=7$ periodicity in the Janus frame is typically 10-15 km, similar to the $12.8$ km found for the occultation data, but in several cases the amplitude approaches $20$ km.
In addition to the expected dominance of $m=6$ in the local frame, we also find evidence for an  $m \simeq 4$ signature in many of the image sequences, with amplitudes of 5-10 km.


\begin{figure}[!h]
	\center\includegraphics[angle=0,width=0.6\columnwidth]{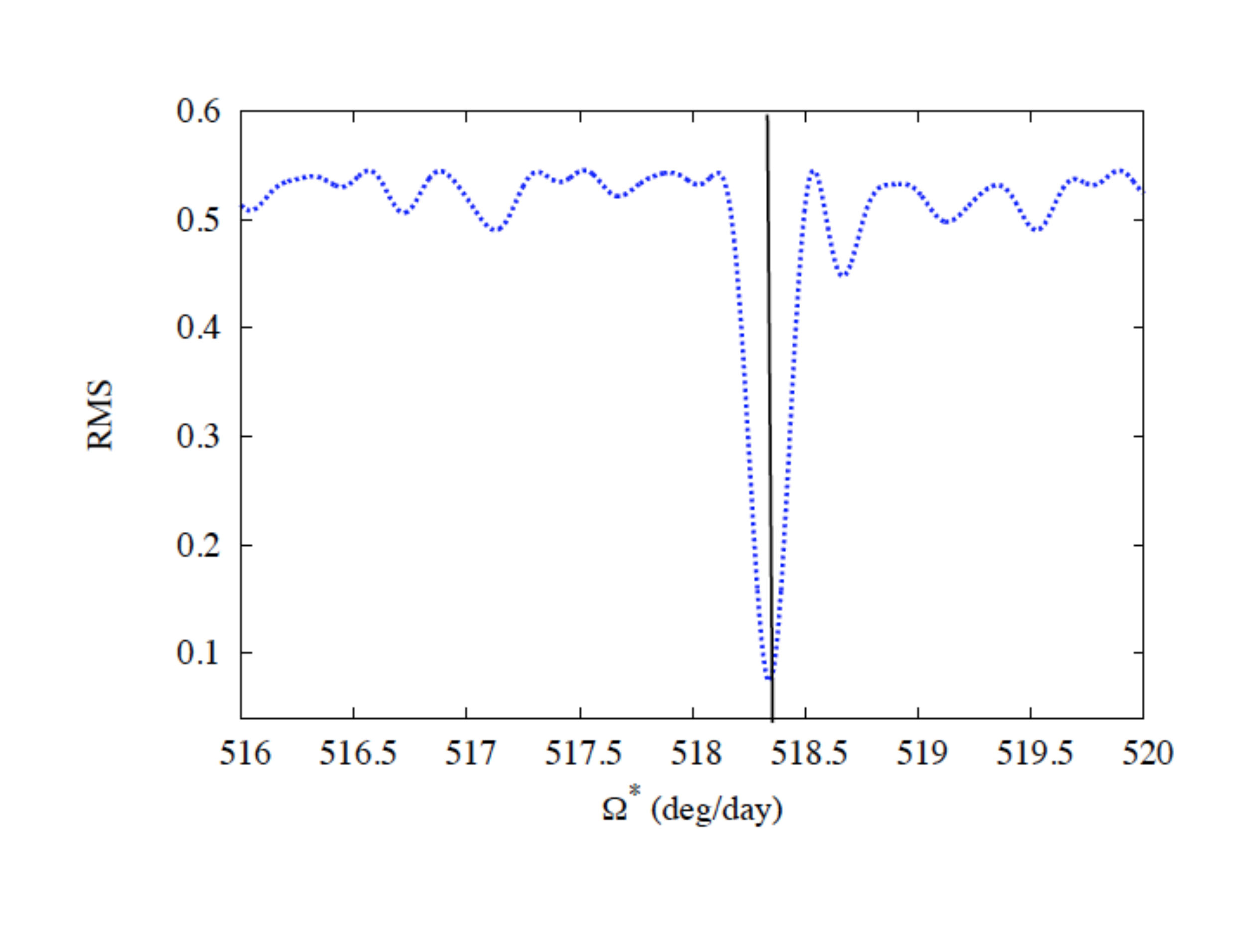}
\centerline{} 
\caption{%
A plot illustrating the presence of a coherent $m=7$ perturbation on the outer edge of the A ring corresponding to $\Omega_p = 518.36 \pm 0.1 \dd$, using imaging sequences obtained between 2006 and 2009.
The RMS phase difference $\Psi$ is calculated using Eq. (\ref{rms_psi}) for an assumed value of $m=7$. A vertical line indicates the predicted pattern speed, $\Omega_p=n_J$.
}%
\label{rms_518}
\end{figure}

In order to verify the role of the 7:6 Janus ILR in the pre-2010 data, we use Eq. (\ref{rms_psi}) with $m=7$
to calculate the RMS variation of the phase $\Psi$, between movies. 
The results are shown in Fig. \ref{rms_518} where we see a deep minimum at $\Omega_p = 518.36 \pm 0.1\dd$, coinciding with the mean motion of Janus in this period.

\begin{figure}[!h]
	\center\includegraphics[angle=0,width=0.6\columnwidth]{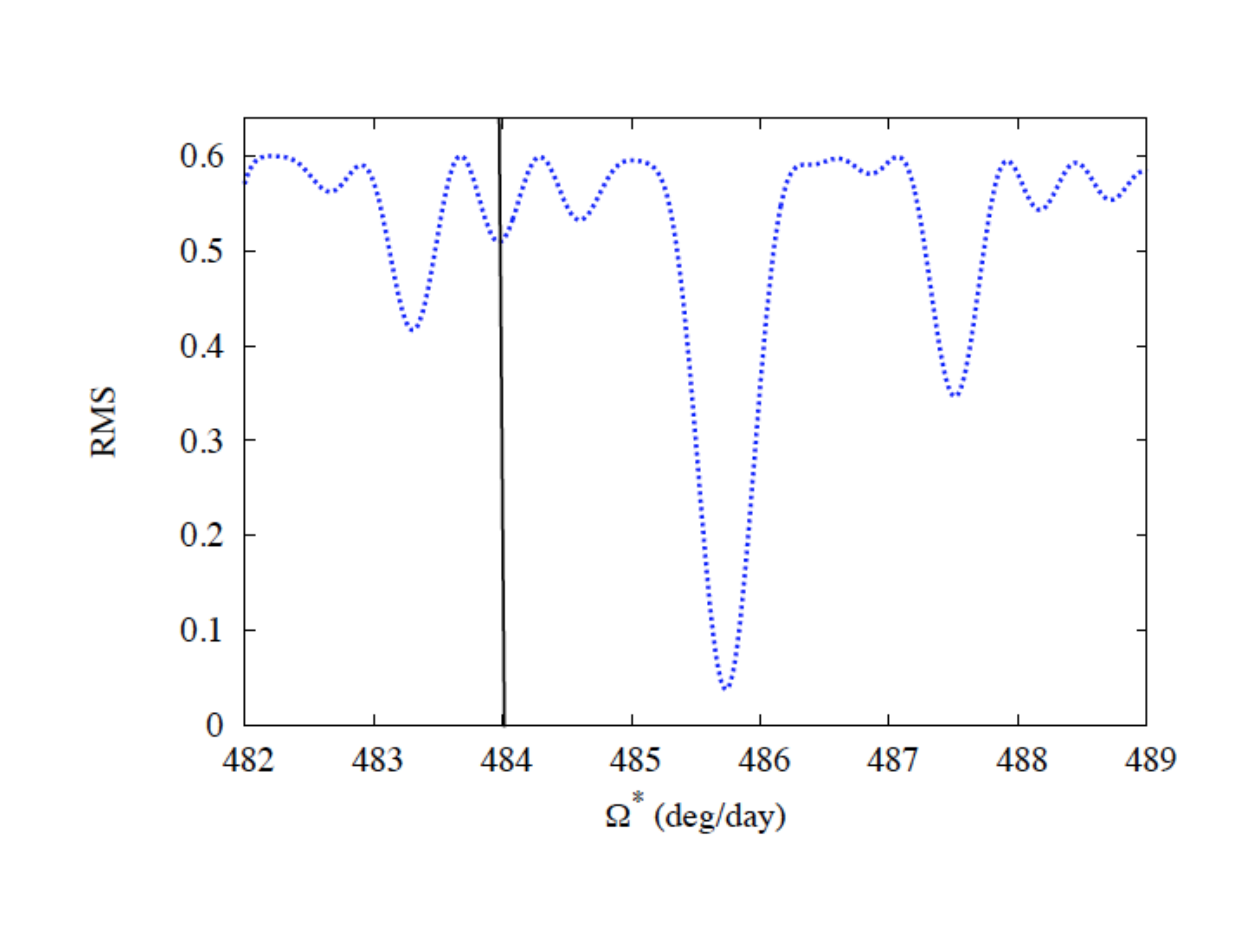}
\centerline{} 
\caption{%
A plot illustrating the possible presence of a coherent $m=5$ ILR-type mode on the outer edge of the A ring in the period between 2006 and 2009, corresponding to $\Omega_p = 485.75 \pm 0.25\dd$. A vertical line indicates the predicted pattern speed, $483.96\dd$.
}%
\label{rms_m_5}
\end{figure}

In order to compare with the occultation data results, we next carried out a similar analysis for $m=5$, with the results shown in Fig. \ref{rms_m_5}. We found a strong minimum at $\Omega_p = 485.75 \pm 0.25\dd$, but the difference of $1.7\dd$ from that seen in the pre-2010 occultation data (cf. Fig \ref{occdata_m5_before}) has no obvious explanation. 

Although the concentration of periodic signals with $m \simeq 4$ in Fig. \ref{m_amp_before_edge} in the local frame (i.e., $\Omega^{*} = n$) may indeed be due to such an $m=5$ ILR-type normal mode, it is also possible that it reflects the existence of an $m=3$ OLR-type mode, as pointed out in Section 2.
This suggestion is motivated by the observation that the 3:4 tesseral resonance with Saturn itself (which is an OLR with $m=3$ and $\Omega_p \simeq \Omega_{Saturn} $) lies close to the edge of the A ring \citep{hedman09}.

\begin{figure}[!h]
	\center\includegraphics[angle=0,width=0.6\columnwidth]{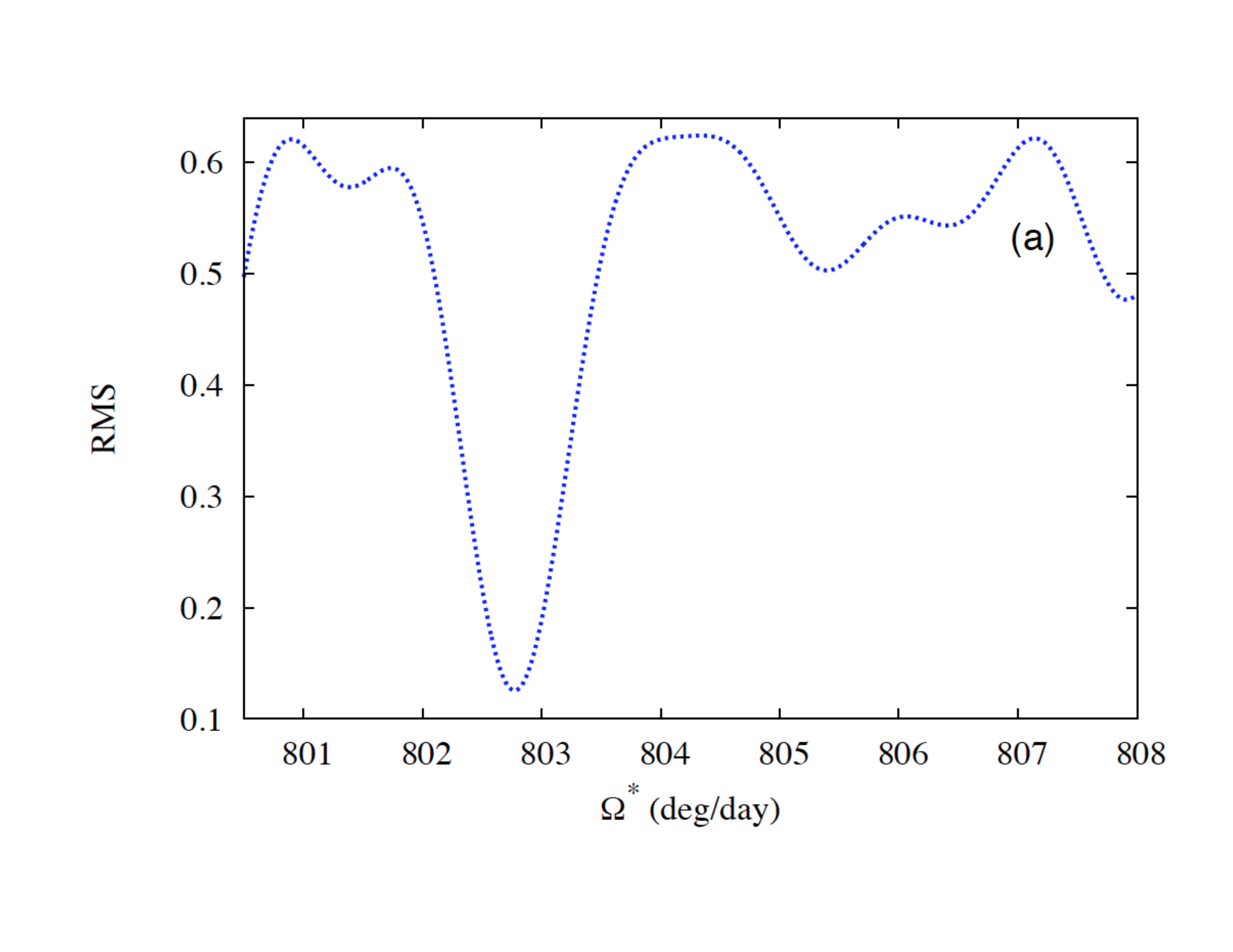}
\centerline{} 
\caption{%
A plot illustrating the possible presence of a coherent $m=3$ OLR-type mode on the outer edge of the A ring in the period between 2006 and 2009 corresponding to $\Omega_p = 802.76 \pm 0.5\dd$.
}%
\label{rms_m_3}
\end{figure}

A scan for such an $m=3$ OLR-type mode is shown in Fig. \ref{rms_m_3}, where we find a strong minimum at $\Omega_p = 802.76 \pm 0.5\dd$, though it is not so deep or narrow as that for $m=5$ in Fig. \ref{rms_m_5}.
This is within the range of rotation rates inferred from Saturnian radio emissions \citep{gurnett07,lamy11}.

In this section, we have confirmed the strong influence of the 7:6 ILR  when Janus was in its inner orbital position between 2006 and 2009.
In addition to this result, we have identified an ILR-type normal mode with $m=5$ in both occultation and imaging data.
Moreover, we find in Figs. \ref{m_amp_before_edge} and \ref{m_amp_before_janus} that, besides the 7:6 signature, we have a persistent $m \simeq 4$ pattern in the keplerian frame.
This could be a signature corresponding to a 3:4 OLR, with $m=4$ in the local frame (or $m=3$ in the Saturnian frame), 
where the pattern speed corresponding to this harmonic is close to the rotation rate of Saturn.

However, we acknowledge some difficulties with this interpretation. On the observational side, this signal seems absent in the occultation data during the same time interval, whereas on the theoretical side, we do not expect an OLR-type mode at an outer ring edge. At this time, we do not see a way to reconcile these viewpoints.

\subsection{Analysis of the data taken between 2010 and 2014}

We now address the question of whether these patterns or others are present in the behavior of the A-ring's outer edge when Janus is orbiting in its outer position, such as it was between 2010 and 2014.

\subsubsection{Occultation data}
 
The period between June 2010 and July 2013 yielded a total of only 35 occultation profiles. This relatively small number reflects the fact that Cassini spent most of 
2010 and all of 2011 in equatorial orbits, where ring occultations are difficult, if not impossible. Our data for this period thus come mostly from the period between July 2012 and 2013, with just two
occultations in mid-2010.  Relative to a circular ring edge, the RMS scatter in the measured radii is
8.4~km, slightly less than that seen before 2010 but still substantially larger than our measurement errors or systematic uncertainties of $\sim300$~m.
Perhaps not surprisingly, since the Janus 7:6 ILR has now moved well outside the ring edge,
the data show no evidence for a coherent $m=7$ signature in this period, at least with a pattern speed close to that of Janus' mean motion. 
 
\begin{figure}[!h]
\centerline{\includegraphics[angle=0,totalheight=15cm,trim= 0 0 0 0]{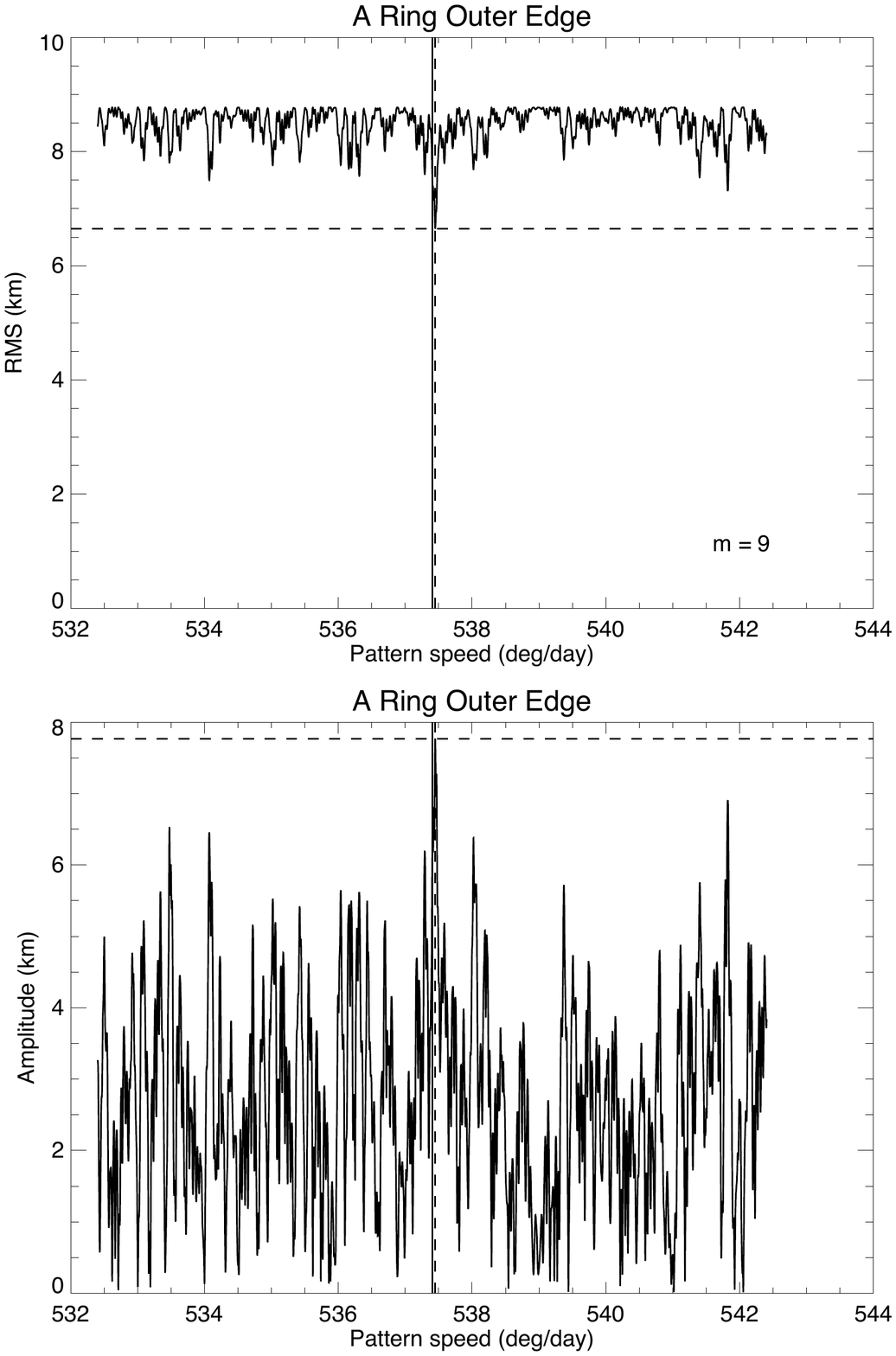}} 
\caption{
Identification of an $m=9$ normal mode on the outer  edge of the A ring in occultation data obtained between June 2010 and July 2013. The format is the same
as that of Fig.~\ref{occdata_m7_before}. The vertical dashed line indicates the rate that provides the best fit to the data (in this case
$\Omega_p = 537.450\dd$), while the solid line indicates the predicted rate for an $m=9$ ILR which falls at the fitted semimajor axis of the edge, or $537.403\dd$.
The maximum amplitude is 7.9~km.}
\label{occdata_m9_after}
\end{figure}

\begin{figure}[!h]
\centerline{\includegraphics[angle=0,totalheight=15cm,trim= 0 0 0 0]{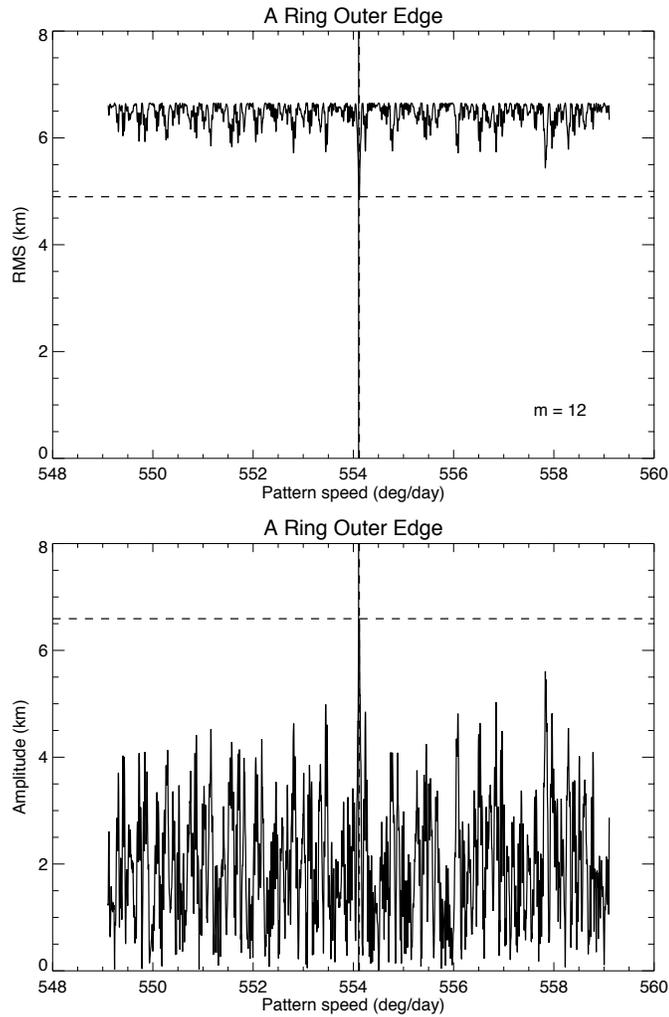}} 
\caption{
Identification of an $m=12$ normal mode on the outer  edge of the A ring in occultation data obtained between June 2010 and July 2013. The format is the same
as that of Fig.~\ref{occdata_m7_before}. The vertical dashed line indicates the rate that provides the best fit to the data (in this case
$\Omega_p = 554.119\dd$), while the solid line indicates the predicted rate for an $m=12$ ILR which falls at the fitted semimajor axis of the edge, or $554.104\dd$.
The maximum amplitude is 6.6~km. For this fit the best-fitting $m=9$ model from  Fig.~\ref{occdata_m9_after} was first subtracted from the raw data.}
\label{occdata_m12_after}
\end{figure}

However, our spectral-scanning program revealed the existence of two fairly strong ILR-type modes, one with 
$m=9$ and the other with $m=12$ (see Figs.~\ref{occdata_m9_after} and \ref{occdata_m12_after}). Least-squares fits again confirm the reality of both modes, with
pattern speeds of $\Omega_p = 537.450\pm0.008\dd$ and $\Omega_p = 554.119\pm0.005\dd$, respectively, and amplitudes of $A_m = 6.1\pm1.2$~km and $7.9\pm1.4$~km. Complete sets of fit
parameters are again given in Table 8.

Unlike the situation prior to 2010, we find no evidence for the existence of additional, weaker ILR-type normal modes in this period.
Our adopted best fit thus consists of only the $m=9$ and $m=12$ modes and has a post-fit RMS residual of 4.3~km, significantly greater than
that obtained for the pre-2010 data.  Again, we find no evidence for OLR-type modes ({\it i.e.,} modes with $m<0$).

\begin{figure}[!h]
\centerline{\includegraphics[angle=0,width=0.55\columnwidth]{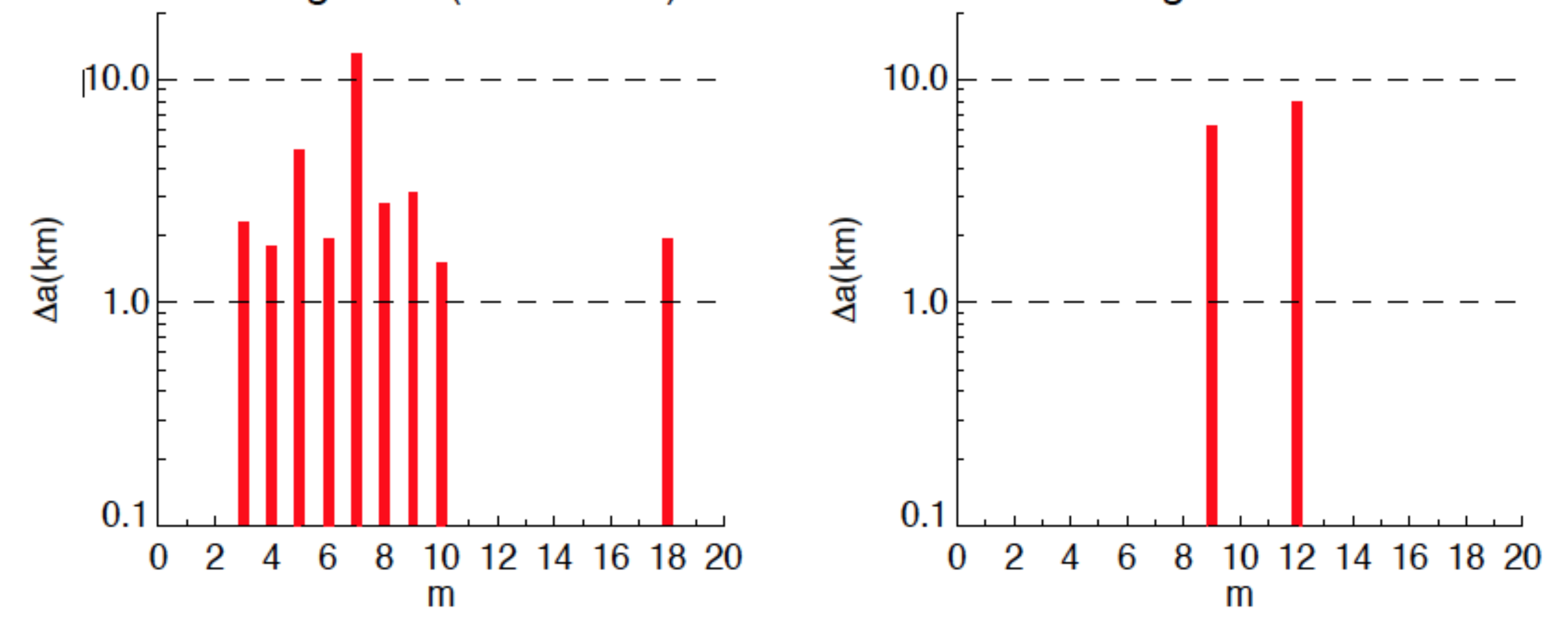}} 
\caption{
Histograms of the fitted amplitudes for ILR-type modes identified in the occultation data for the periods 2006-2009 and 2010-2013,
corresponding to the two legs of the coorbital satellite libration. Note the dominance of the $m=7$ mode in the first period, due to
the Janus 7:6 ILR.}
\label{occhistograms}
\end{figure}

Fig.~\ref{occhistograms} summarizes the amplitudes of all ILR-type modes identified in the occultation data for the periods 2006--2009 and
after 2010.

\subsubsection{Imaging data}

We have carried out a similar analysis for the 9 imaging sequences taken between 2010 and 2013 as described in Section 4.1. 
Again because Cassini was on equatorial orbits in 2010 and 2011, all of the available data come from 2012 and 2013 (see Table 1).
Frequency analysis confirms that the effect of the 7:6 Janus ILR disappears in this period after to 2010.
However, other perturbations still affect the edge, 
including the $m=4$ signature noted above in the `local' frame.

\begin{figure}[!h]
	\center\includegraphics[angle=0,width=0.75\columnwidth]{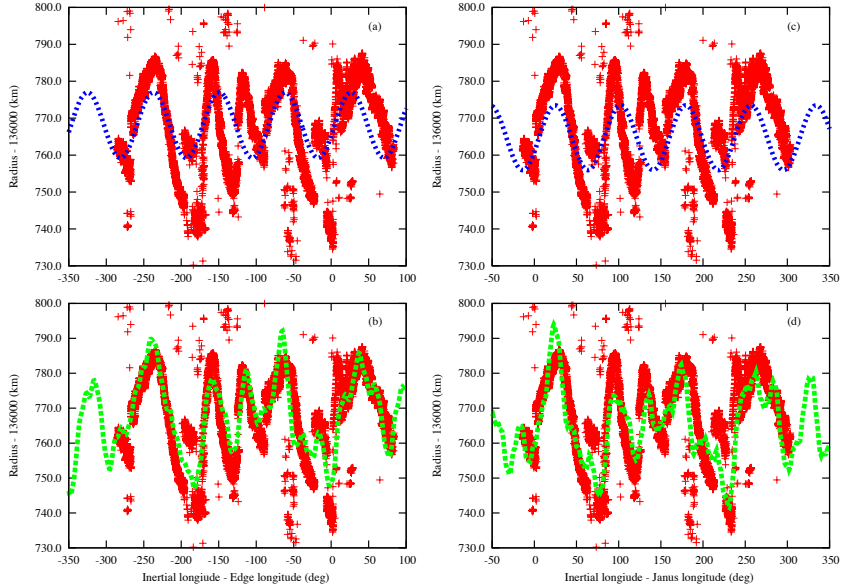}
\caption{%
Measured radii from Cassini images for movie number 29 (see Table \ref{tab_movie_iss}) taken in May 2013,
mosaicked in the keplerian frame (left panels) and in the Janus frame (right panels).
Panels (a) and (c) show the dominant frequency in each frame (green curve), while panels (b) and (d) show the combined signal of the larger components (blue curve). Note that points are the data and the curves are the least-squares models.
}%
\label{famous_after}
\end{figure}

Specifically, when mosaics are made using $\Omega^{*} = n_J = 518.24 \dd$, the $m=7$ pattern seen prior to 2010 is absent, 
and, if the data are instead mosaicked with  $\Omega^{*} = n = 604.22 \dd$ we do not see the familiar 6-lobed pattern.
This behavior is illustrated in Fig. \ref{famous_after} for movie $29$ with least-squares fit parameters given in Tables \ref{tab_famous_edge_after} and \ref{tab_famous_janus_after}.
As in Fig. \ref{famous_before}, panels (a) and (b) show the measured radii of the A ring edge 
for $\Omega^{*} = n$, 
while panels (c) and (d) show the same measurements, 
but mosaicked with $\Omega^{*} = n_J$.
 
Although the range between radii seen in Fig. \ref{famous_after} is similar to that in Fig. \ref{famous_before}, or $\sim 50$ km peak-to-peak, the variation with longitude appears chaotic rather than quasi-periodic. 
A similar impression is gained by comparing the images themselves (Figs. \ref{mos_before} and \ref{mos_after}) above.

%
%


%
\begin{table}[H]
\vskip4mm
\centering
\begin{tabular}{|*{5}{c|}}
\hline
  Frequencies & Amplitudes (km) & Period (deg) & Phase (deg)  &  $m$  \\
\hline
 1 & 8.87 & 87.89 & 251.15 & 4.09  \\
\hline
 2 & 8.36 & 155.45 & 218.38 & 2.31  \\
\hline
 3 & 5.53 & 42.81 & 215.87 & 8.40  \\
\hline
\end{tabular}\\
\caption{Frequency analysis for the data in Figure \ref{famous_after} with $\Omega^{*} =n$.}
\label{tab_famous_edge_after}
\end{table}
\begin{table}[H]
\vskip4mm
\centering
\begin{tabular}{|*{5}{c|}}
\hline
  Frequencies & Amplitudes (km) & Period (deg) & Phase (deg)  &  $m$  \\
\hline
 1 & 8.76 & 77.29 & 243.88 &  4.66  \\
\hline
 2 & 7.99 & 131.50 & 309.52 & 2.73 \\
\hline
 3 & 6.39 & 37.67 & 144.33 & 9.55 \\
\hline
\end{tabular}\\
\caption{Frequency analysis for the data in Figure \ref{famous_after} with $\Omega^{*} =n_J$.}
\label{tab_famous_janus_after}
\end{table}

\begin{figure}[!h]
	\center\includegraphics[angle=0,width=0.6\columnwidth]{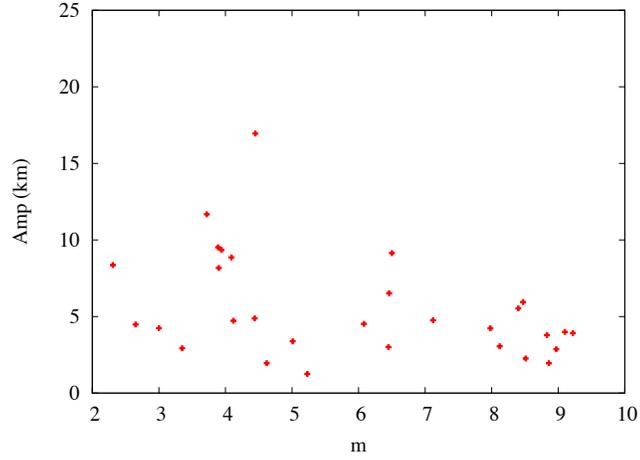}
\centerline{} 
\caption{%
Results of frequency analysis for imaging sequences obtained between 2012 and 2013 using $\Omega^{*} = n$.
Note the preference for $m \simeq 4$ in the local frame.
}%
\label{m_amp_after_edge}
\end{figure}

\begin{figure}[!h]
	\center\includegraphics[angle=0,width=0.6\columnwidth]{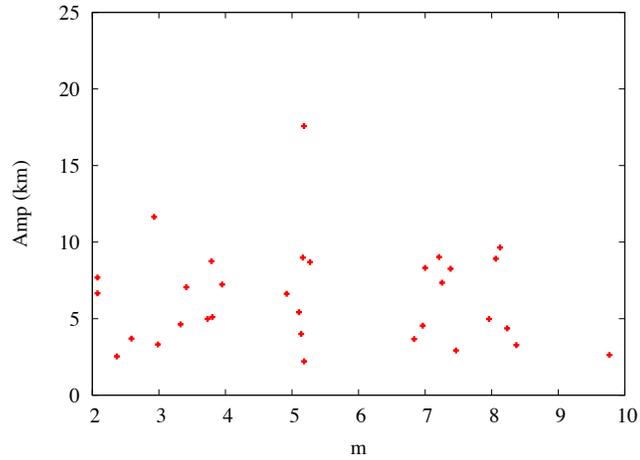}
\centerline{} 
\caption{%
Results of frequency analysis for imaging sequences obtained between 2012 and 2013 using $\Omega^{*} = n_J$.
Note the absence of a significant signal at $m=7$ in the Janus frame.
}%
\label{m_amp_after_janus}
\end{figure}

Fig. \ref{m_amp_after_edge} (in the local frame) and Fig. \ref{m_amp_after_janus} (in Janus frame) show the amplitude of the fitted edge variations 
for all of the image sequences in 2012-2013 versus $m$, for $\Omega^{*} = n$ and $\Omega^{*} = n_J$, respectively.
The fit parameters are summarized in Table \ref{table_famous_edge_after} in Appendix A.
Note the dominance of $m \simeq 4$ perturbations in the local frame and the disappearance of the $m=7$ pattern in the Janus frame. 

\begin{figure}[!h]
	\centerline{\includegraphics[angle=0,width=0.6\columnwidth]{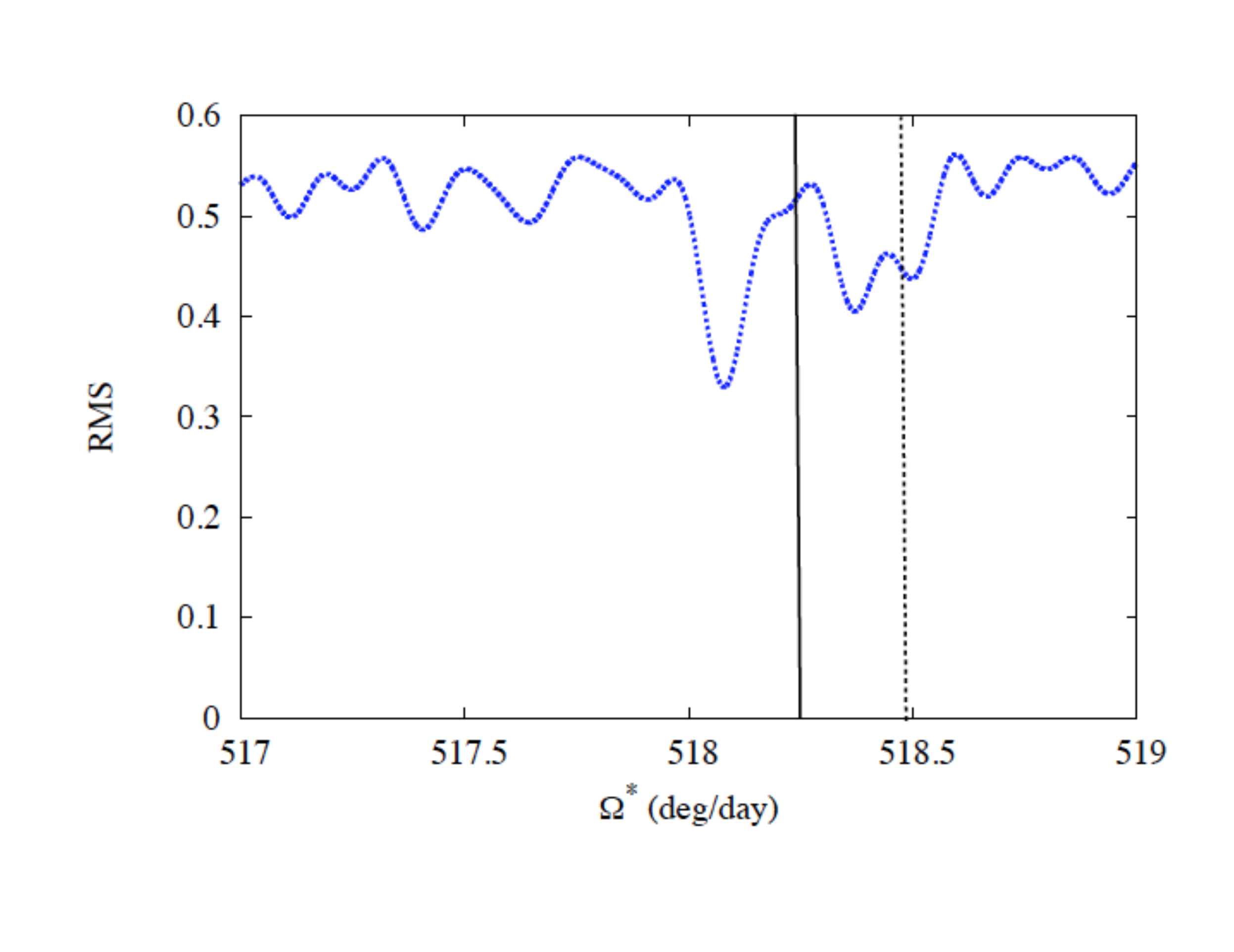}} 
\caption{%
A plot illustrating the absence in 2010-2013 of a coherent $m=7$ perturbation on the outer edge of the A ring, which we would have expected to see at $\Omega_p = n_J = 518.24 \dd$, as indicated by the solid vertical line. The dotted line corresponds to Epimetheus' pattern speed at this period.
The RMS phase difference is calculated using Eq. (\ref{rms_psi}) for an assumed value of $m=7$.
}%
\label{no_518}
\end{figure}

In order to verify the absence of the 7:6 Janus ILR signature in the post-2010 period, we again calculate the RMS variations of the phase using Eq. (\ref{rms_psi}), with results shown in Fig. \ref{no_518}.
We see only a weak minimum at $\sim 518.1 \pm 0.05\dd$, and nothing at the expected frequency of $n_J \simeq 518.24 \dd$ for this period.
Also, we do not observe any clear signature from Epimetheus, which we would expect to see at a pattern speed of $\Omega_E=518.49\dd$.

\begin{figure}[!h]
	\center\includegraphics[angle=0,width=0.6\columnwidth]{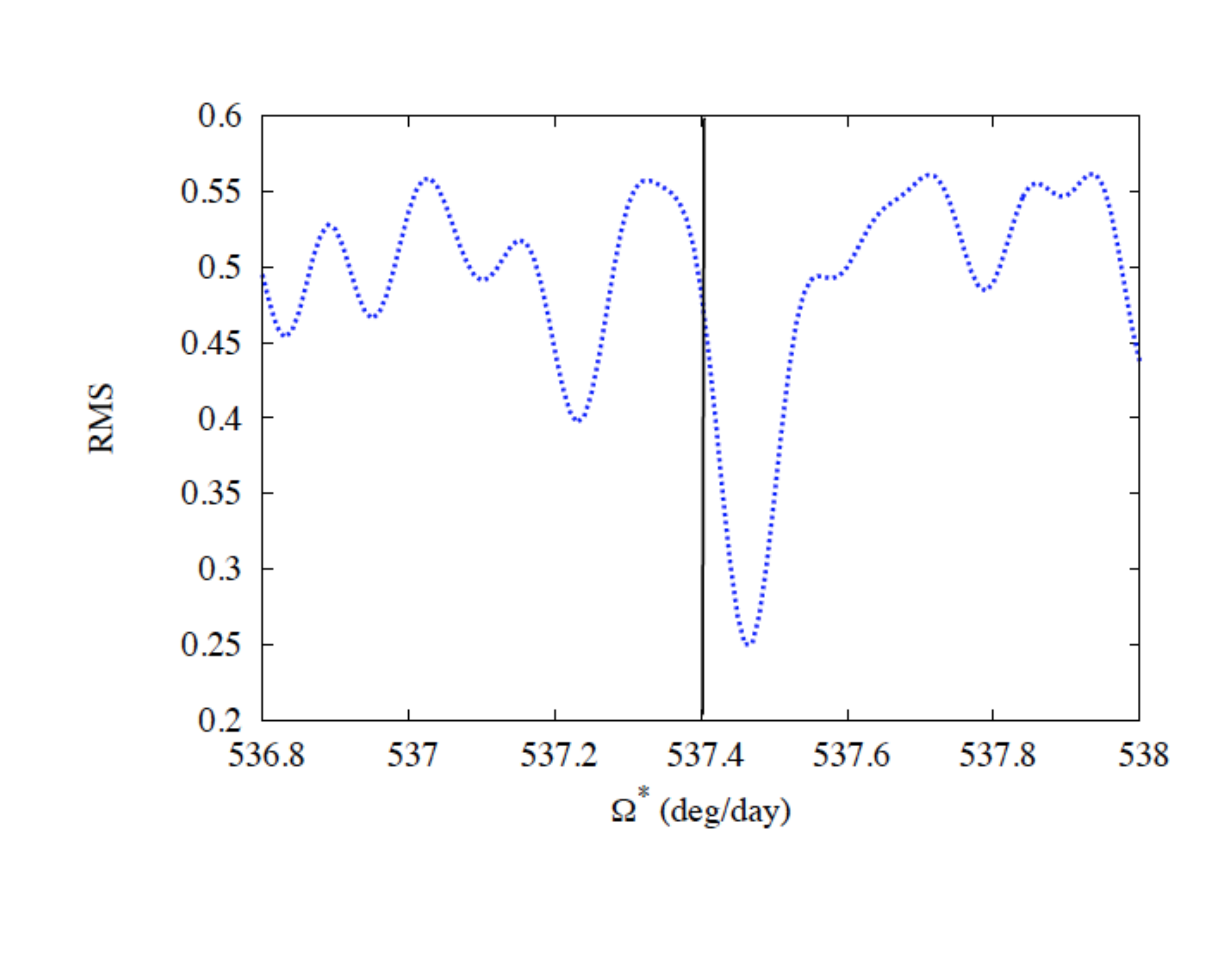}
\centerline{} 
\caption{%
A plot illustrating the presence of an $m=9$ ILR-type mode in 2012-2013 on the outer edge of the A ring corresponding to $\Omega_p = 537.45 \pm 0.04\dd$ for the time period from 2009 and 2013. A vertical line indicates the predicted pattern speed, $537.40\dd$. 
}%
\label{rms_9}
\end{figure}

\begin{figure}[!h]
	\center\includegraphics[angle=0,width=0.6\columnwidth]{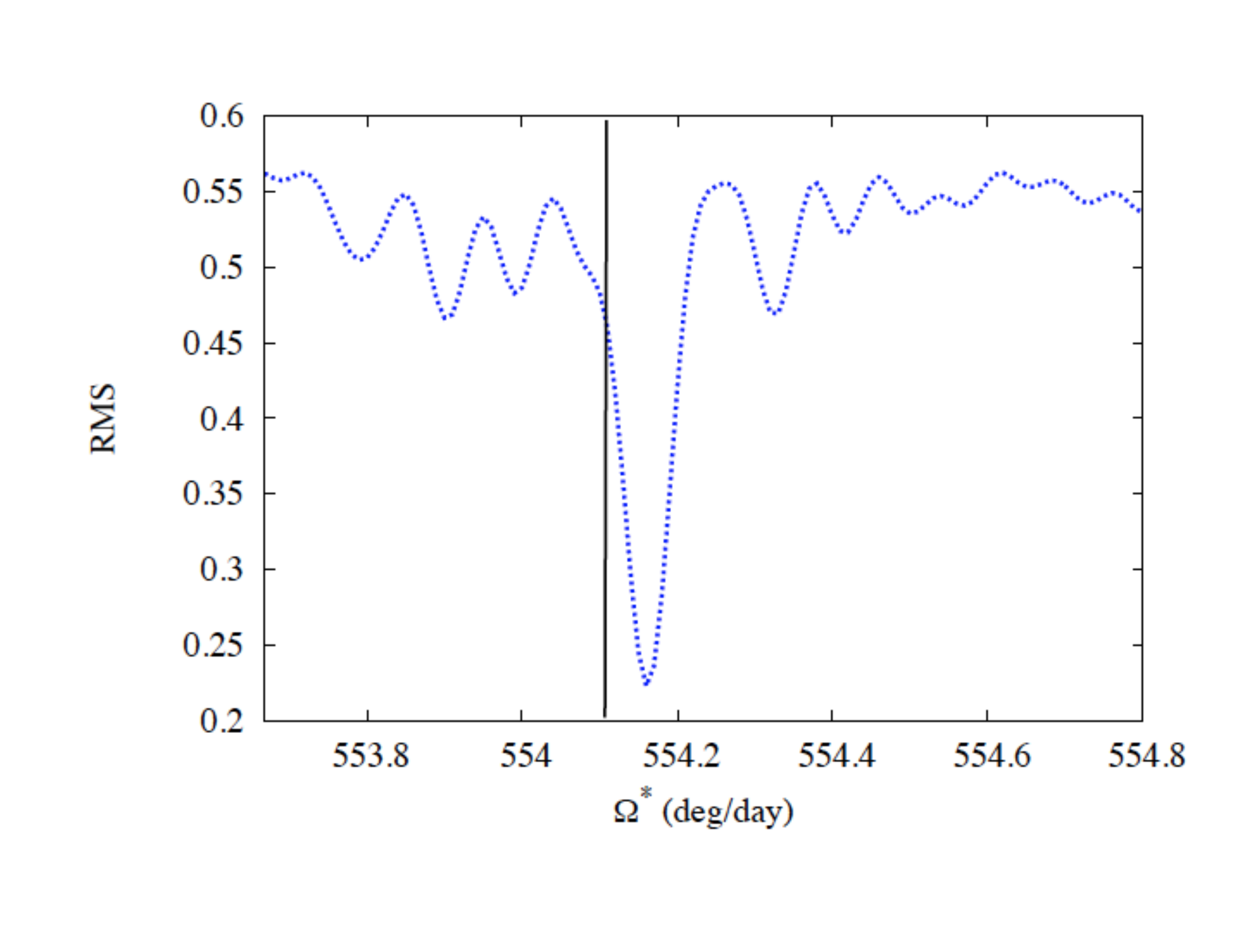}
\centerline{} 
\caption{%
A plot illustrating the presence of an $m=12$ ILR-type mode in 2012-2013 on the outer edge of the A ring corresponding to $\Omega_p = 554.15 \pm 0.04\dd$ for the time period from 2009 and 2013. A vertical line indicates the predicted pattern speed, $554.10\dd$. 
}%
\label{rms_12}
\end{figure}

As the occultation data for 2010-2013 period instead show ILR-type modes with $m=9$ and $m=12$, 
we have also searched for these signals in the imaging data, as shown in Figs. \ref{rms_9} and \ref{rms_12}.
We find moderately deep minima in both cases, of $537.45 \pm 0.04\dd$ and $554.15 \pm 0.04\dd$, respectively, which agree well with the pattern speeds seen in the occultation data, 
as well as with the expected values for $m=9$ and $m=12$ ILR-type normal modes.

\begin{figure}[!h]
	\center\includegraphics[angle=0,width=0.6\columnwidth]{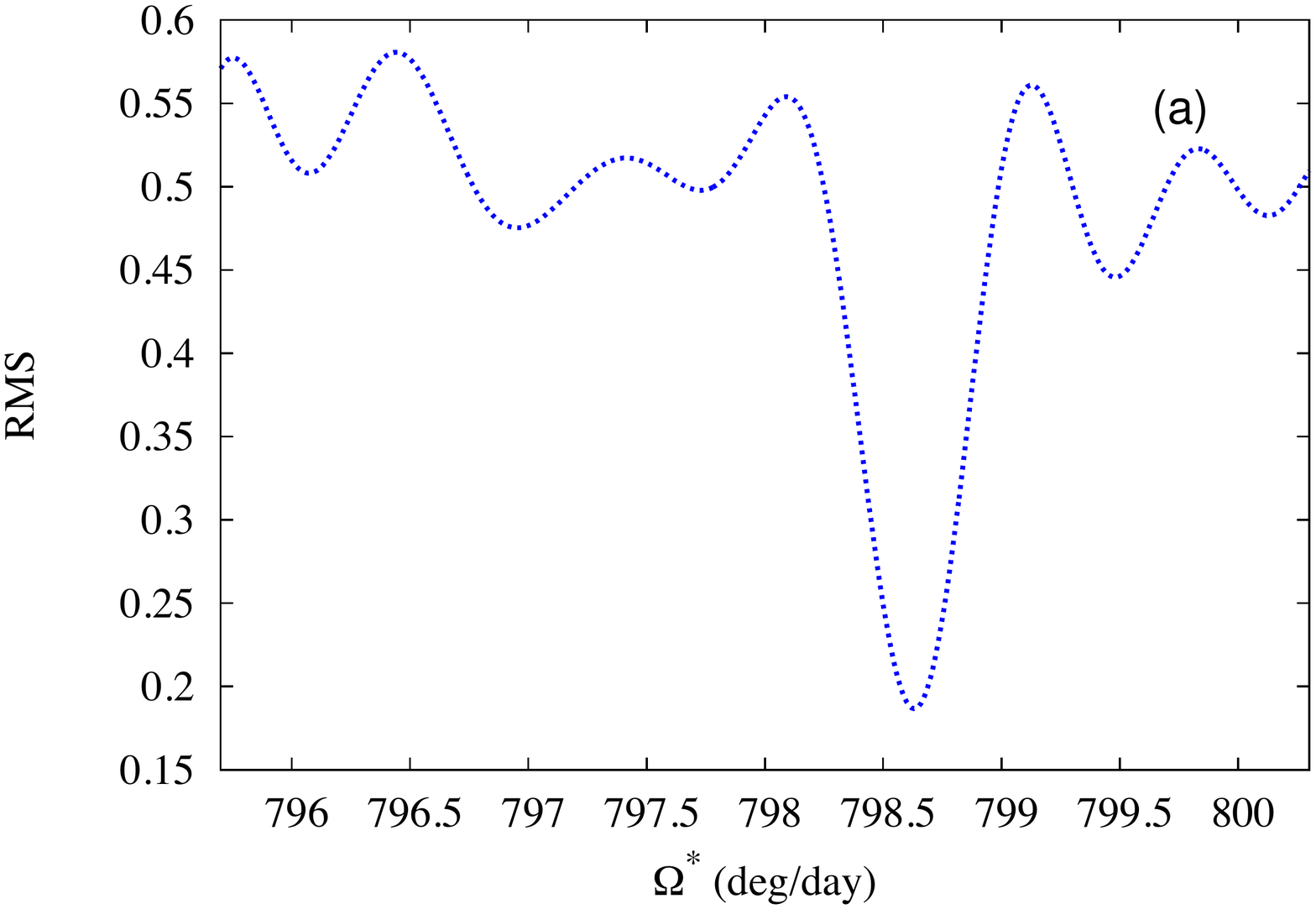}	
\centerline{} 
\caption{%
A plot illustrating the possible presence of an $m=3$ OLR-type mode in 2012-2013 on the outer edge of the A ring corresponding to $\Omega_p = 798.65 \pm 0.25\dd$. 
}%
\label{rms_3_after}
\end{figure}

Finally, we checked again for an $m=3$ OLR or an $m=5$ ILR as potential explanations for the concentration of points near $m=4$ seen in Fig. \ref{m_amp_after_edge}. 
Fig. \ref{rms_3_after} shows a possible $m=3$ OLR-type mode, similar to what we found in Fig. \ref{rms_m_3} for the data prior to 2010, although no mode of this period was found in the occultation data. No evidence was found for an $m=5$ ILR-type mode.

In this subsection, we have confirmed that 2 years after Janus moved outward, the ring is no longer appreciably influenced by the 7:6 Janus ILR.
However, we do find normal modes for $m=9$ and $m=12$ in both occultation and imaging data sets.
Moreover, as in the period prior to 2010, we also find evidence in the imaging data for an OLR-type mode with $m=3$.
However, we note that  the minimum RMS in Fig. \ref{rms_3_after} corresponds to $\Omega_p = 798.65 \pm 0.25\dd$, which is significantly slower than the $m=3$ pattern speed seen prior to 2010 in Fig. \ref{rms_m_3}.

\section{Interpretations and discussions}

\subsection{Perturbations due to the Janus 7:6 resonance}

It is no surprise that, prior to January 2010,  the radial perturbations of the A ring's outer edge are dominated by
the 7:6 ILR with Janus. During this period, this strong resonance was located within a few km of the mean radius
of the edge (cf. Fig.~\ref{orbit_swap}). 
Our results are consistent with those of
\cite{spitale09}, who found a strong $m=7$ resonant signature in their analysis of 
Cassini imaging sequences obtained between 2006 and 2009. 
Both studies found similar amplitudes
of 12-15~km, and pattern speeds consistent with Janus' mean motion on the inner leg of its coorbital motion, i.e., $\Omega_p=518.354\dd$.

However, 
when Janus is on the outer leg of its co-orbital motion, the behavior of the edge is unclear. \cite{spitale09} found
the ring edge to be perturbed in the interval a few months prior to the orbital swap in January 2006, but with a smaller amplitude and in an apparently 
disorganized manner. They suggested that this situation reflected the onset of the swap, and noted also that 
a stable $m=7$ pattern only established itself about 8 months after the orbital exchange was complete. 
Our data, both from occultations and imaging sequences, show a complete absence of any recognizable
$m=7$ signature between July 2012 and the end of 2013, but unfortunately we have no suitable data taken immediately after the coorbital swap in January 2010. 
Interestingly, this interval (when the Janus 7:6 ILR was located $\sim 15$ km exterior to the A ring's outer edge) corresponds to the same phase of the coorbital libration when the Voyager encounters occurred, which raises the question of how \cite{porco84} were able to identify an $m=7$ signature in their data.
Either the resonant forcing was stronger at that time, or the very limited number of Voyager measurements (a total of ten, nine of which were made over a period of a few days by Voyager 2)
may have led to a misidentification of the nature of the perturbations. 
In this context, we note that 
(i) the \cite{porco84} amplitude is only about one-half that seen in 2006-2009;
(ii) the limited temporal sampling of the Voyager data permitted multiple solutions, with pattern speeds differing by integer multiples
of $0.18\dd$ (see their Fig. \ref{orbit_swap}), from which they chose the value that most closely matched 
the coorbitals' mass-weighted average mean motion of $518.29\dd$; and
(iii) the Voyager data actually sampled only four of the putative 7 lobes (see their Fig. \ref{Cassini_image_1615}).
In hindsight, the near-coincidence between the strong Janus  7:6 ILR and the outer edge of the A ring may have unduly limited the range of models they considered.

In fairness, we should also note that \cite{porco84} suggested a very different way to visualize and model the coorbital satellite's 
ILRs than that adopted here (as illustrated in Fig. 2 above) and by \cite{spitale09}.
In our picture, each satellite generates its own 7:6 ILR, separated by $43$ km from one another, and these two resonances 
shift abruptly in radius and pattern speed when the satellites exchange orbits every $4$ years and Janus' mean motion alternates between $518.238 \dd$ and $518.345\dd$ \citep{jacobson08}.
In the model adopted by \cite{porco84}, this explicit time dependence is dropped and the two discrete ILRs are replaced by a series 
of closely-spaced sub-resonances centered on the mass-weighted average mean motion of $518.2918 \dd$.
The resonance spacing is $\Omega_L/7 \simeq 0.0176\dd$, where $2\pi/\Omega_L$ is the coorbital libration period of $8$ years.
\cite{porco84} noted that four of these sub-resonances fell within the $\pm1 \sigma$ error bounds of their best-fitting pattern speed of $518.31 \pm 0.04 \dd$.
In this scenario, there is no obvious reason for the 7-lobed perturbation at the edge of the A ring to change either its amplitude or pattern speed at each 
orbital swap.
However, our observations are inconsistent with this picture, and support instead the model described in Fig. \ref{orbit_swap}.

But the ring is certainly not unperturbed in the period when Janus is in its outer orbital position and its ILR is located well outside the ring edge.  
The occultation and imaging data after January 2010 instead show strong evidence for ILR-type normal modes with $m=9$ and 12. 
Both before and after 2010, the imaging data also consistently
show evidence for an $m=4$ perturbation in the keplerian frame. 
As noted in section 4.1 above, the latter could reflect either an ILR-type mode with
$m=5$, and/or an OLR-type mode with $m=-3$\footnote{The negative m's represents the OLR and OLR-type mode}. 
In the next two subsections, we will  investigate possible explanations for these unexpected perturbations.

\subsection{Normal ILR-modes and OLR-modes}

With the exception of the $m=7$ mode, which is undoubtedly forced by the Janus 7:6 resonance, and the OLR-type
$m=-3$ mode seen in the imaging data, all of the remaining perturbations  are consistent with ILR-type normal modes
spontaneously generated at the edge of the ring. As described by \cite{nicholson14b}, such modes can be thought of
as arising when an outward-propagating trailing spiral  density wave, driven at a resonant location just interior to the
ring edge, reflects at the sharp outer edge to form an inward-propagating leading wave. The leading wave, in turn,
reflects at the Lindblad resonance to reinforce the outward-propagating trailing wave, resulting in a standing wave that  
exists only in the narrow range between the resonance and the ring edge. For a given value of $m$, the pattern speed
of such a standing wave is given by Eq.~(\ref{patspeed}), where the mean motion and apsidal precession rate are evaluated 
at the Lindblad resonance.

We can test this interpretation by using the observed value of $\Omega_p$ for each mode to calculate the corresponding
resonance location, $\ares$, and thus the offset $\Delta a_p = \ares - a_0$, where $a_0 \simeq$ 136,770 km is the 
mean radius of the A-ring edge.  For each fitted mode, the last column of Table 8 lists the calculated value of $\Delta a_p$.
We find that, as expected, all values are negative, ranging from $-13$~km for $m=3$ to $-4$~km for $m=12$.  Although by
no means perfect, there is also a general trend for $\Delta a_p$ to decrease in magnitude as $m$ increases, in agreement
with theoretical expectations for the resonant-cavity model \citep{spitale10,nicholson14a}.
We note also that, as density waves of a given pattern speed and m-value propagate only in the regions exterior to ILR and interior to OLR, 
only  ILR-type normal modes are expected to be found at the outer edge of a broad ring, in agreement with the occultation data. 

From Eq. (19) of \cite{nicholson14a}, which is based on the dispersion relation for density waves, we can use the calculated values of $\Delta a_p$ to estimate the surface density of the outermost part of the A ring.
We find values ranging from $2.4$~$g.cm^{-2}$ (m=2) to $12.9$~$g.cm^{-2}$ (m=5), with an average of $8.4$~$g.cm^{-2}$.
This is significantly less than typical A ring values of $30-40$~$g.cm^{-2}$ \citep{tiscareno07}, but might reflect the increasing fraction of small particles in this region \citep{cuzzi09}. Given the mass densities are so low, the dynamics of these sorts of normal modes are still unclear and might not yield robust mass estimates.

\subsection{Saturn Tesseral resonances}

After the $m=7$ perturbation due to the Janus resonance, the most persistent periodic signature in the imaging data for the interval 2006-2009 is an $m=4$ mode in the local
or keplerian frame (see Table \ref{tab_famous_edge_before} and Fig. \ref{m_amp_before_edge}).
As noted in Section 4.1, this could represent either an $m=5$ ILR-type mode or an $m=-3$ OLR-type mode, following Eq. (\ref{radius3}).
Phase analysis of these data, shown in Figs. \ref{rms_m_5} and \ref{rms_m_3} suggests that indeed both modes are present
though the pattern speed for $m=5$ differs significantly from the predicted value.  
After 2010, we again see evidence for an $m=4$ perturbation in the local frame (see Table \ref{tab_famous_janus_before} and Fig. \ref{m_amp_after_edge}), 
but in this case phase analysis reveals only the $m=-3$ mode (see Fig. \ref{rms_3_after})
albeit with a lower pattern speed.
The occultation data, however, do not confirm the presence of any significant OLR-type modes, including $m=-3$, either prior to or after 2010.
How can these apparently-contradictory results be reconciled, and what might be the source of such a perturbation?  
 
Looking first at the second problem, we note that only ILR-type normal modes are expected at a sharp outer ring edge, such as that of the A ring.
The only known examples of OLR-type normal modes are found at inner ring edges or in very narrow ringlets \citep{nicholson14b}.
Furthermore, all first-order, satellite-driven Lindblad resonances in the rings are ILRs\footnote{Exceptions to this generalization are found for the gap-embedded moonlets Pan and Daphnis, both of which have high-m OLRs in the outer part of the A ring.}, since their pattern speeds are less than ring particle mean motions (cf. Eq. (\ref{patspeed}) above).
There is, however, one potential source of non-axisymmetric gravitational perturbations with $\Omega_P > n$, and that is Saturn itself.
Such resonances are known as tesseral resonances, and they are closely related to satellite-driven Lindblad resonances \citep{hamilton94}.
For a particular component of the planet's gravity field with m-fold axial symmetry, 
the resonance locations are given by Eq. (\ref{patspeed}), which we may rewrite as:
\begin{equation}
	(m \mp 1)n ~~ \pm ~~ \dot{\varpi} = m \Omega_{p},     
\label{strem_m_syn}
\end{equation}
where here the pattern speed $\Omega_p$ is equal to the planet's interior rotation rate.
The upper (lower) sign corresponds to an inner (outer) tesseral resonance, where $n>\Omega_p$ ($n<\Omega_p$).
For Saturn, the synchronous orbit lies in the central B ring, so Inner Tesseral Resonances (ITR) fall in the D, C and inner B rings, and outer tesseral resonances fall in the outer B and A rings.
In fact, the 3:4 Outer Tesseral Resonance (OTR) would coincide with the outer edge of the A ring for $\Omega_p= 804.64\dd$, a value which lies within the range of periods (between $797\dd$ and $817\dd$) observed for Saturn's kilometric radio emission (SKR) \citep{gurnett07}. These are equivalent to Lorentz resonances, where the disturbance arises from the magnetic field \citep{burns85}.

A clue to our first problem, that of the absence of an $m=-3$ signature in the occultation data, is provided by the observation that, while our best fit to the imaging data prior to 2010 is for $\Omega_p = 802.76\dd$ (cf. Fig. \ref{rms_m_3}), the best fit after 2010 is for $\Omega_p = 798.65\dd$ (Fig. \ref{rms_3_after}).
Indeed, if we subdivide the data into even shorter time intervals we find various values of $\Omega_p$ which differ by a few degrees per day.
This suggests that the fundamental driving frequency for this perturbation is either drifting or changing irregularly over the 8-year interval of our observations.
Such a variation could easily result in a non-detection in the occultation data, as these fits presuppose a regular, coherent perturbation extending throughout the fitted period; any variations in $\Omega_p$ greater than $(1-2)\dd$ would result in scrambling the phase of the signal in the typical interval of several months between occultations. 

Supporting evidence for other tesseral resonances in Saturn's rings has been reported in previous studies of Cassini imaging and occultation data.
In their examination of non-axisymmetric structures in images of the tenuous D and G rings, as well as the Roche division (the region immediately exterior to the A ring), 
\cite{hedman09} found strong evidence for multiple perturbations with rotation periods between $10.52$ and $10.82$ hours ($798.5\dd \leq \Omega_p \leq 820.5\dd$).
In the D ring over a radial range of $\sim1300$ km they identified a whole series of $m=2$ perturbations with the 2:1 ITR (in our notation), while the Roche division showed an $m=3$ perturbation that they attributed to the 3:4 OTR.
Again the observed periods mostly fall within the range of reported SKR periods of $797-817\dd$ \citep{lamy11}, and are close to what is generally assumed to be Saturn's interior rotation rate of $\sim818\dd$ based on minimizing the potential vorticity and dynamics heights \citep{anderson07,read09}.
But because both the D ring and Roche division are dominated by small ($<100~\mu$m) dust grains, and because of the multiplicity of pattern speeds found, \cite{hedman09} attributed the source of the perturbations to electromagnetic interactions with the planet's magnetic field \citep{burns85}, rather than to gravitational perturbations.
More recently, \cite{hedman14} reported the identification of five weak density waves seen in occultation data for the outer C ring with $m=+3$ ILR-type resonances.
Comparison of the density wave phases between stellar occultations up to 300 days apart permitted them to determine accurate pattern speeds for the waves, which fell in the range $807.7\dd \leq \Omega_p \leq 833.6\dd$.
In all likelihood, these five C-ring waves are driven by the 3:2 ITR, but in this case the driving force is clearly gravitational rather than electromagnetic.
If we can apply the same interpretation to the $m=-3$ perturbations seen at the outer edge of the A ring, then both this distortion and the C ring density waves would be driven by the same $m=3$ nonaxisymmetric term in Saturn's gravity field:
in one case we see the 3:2 ITR and the other the 3:4 OTR.
But what then does one make of the wide range of pattern speeds seen for these resonantly-forced structures?
\cite{hedman14} point out (see their Fig. 13) that the values of $\Omega_p$ derived from the density waves, while varying by $>3\%$, all lie within the range of rotation rates reported for Saturn's atmospheric winds and radio emissions.
The same may be said for the possible electromagnetic signatures found by \cite{hedman09}. 

We hypothesize, therefore, that
(i) the strong differential rotation seen at Saturn's cloud tops extends quite deeply into the planet, and 
(ii) there exist local mass or density anomalies in the planet's interior --- perhaps associated with low-wavenumber global waves --- that give rise to gravitational signatures affecting the rings.
At present, there are insufficient data to constrain the depth or magnitude of these anomalies, but \cite{hedman14} estimated an approximate mass of $10^{14}-10^{15}$ kg needed to drive the density waves in the C ring.
A similar argument applied to the A-ring edge waves, which have an amplitude of $\sim 5$ km suggests a mass $\sim1/3$ that of Janus, or $\sim10^{18}$ kg $[\Delta_a/\text{km}]$ where $\Delta_a = a_{res} - a_{edge}$.

In future studies, we plan to examine additional A-ring edge images, in the hope of better characterizing the $m=-3$ perturbations, and to look for evidence of additional tesseral resonances elsewhere in Saturn's main rings.

\section{Conclusions}

Our analysis of 8 years of Cassini imaging and occultation data confirms that, 
for the period between 2006 and 2010, the radial perturbation of the A ring's outer edge is dominated by the 7:6 Janus ILR. 
However, between 2010 and 2014, the Janus/Epimetheus orbit swap moves the Janus 7:6 LR away from the A ring's outer edge, and the 7-lobed pattern disappears.  
Moreover, we have identified a variety of normal modes at the edge of the A ring, with values of ''$m$'' ranging from 3 to 18 and appropriate pattern speeds. These modes may represent waves trapped in resonant cavities at the ring edge (Spitale and Porco 2010, Nicholson et al 2014).
Futhermore, we identified some other signatures, with $m=-3$ consistent with tesseral resonances that might be associated with gravitational inhomogeneities in Saturn's interior. 
One possible explanation for the $m=-3$ mode is the 3:4 outer tesseral resonance, 
which would imply that asymmetries in Saturn's interior are responsible, in part, for the complex structure seen on the outer edge of the A ring.
These signatures may provide information about differential rotation in Saturn's interior.

\section*{Acknowledgements}

We wish to thank the ISS, RSS, UVIS and VIMS Teams for acquiring the data we used in this study.
This work was supported by NASA through the Cassini project.
Carl Murray was supported by the Science and Technology Facilities Council (Grant No. ST/F007566/1) and is grateful to them for financial assistance. Carl Murray is also grateful to the Leverhulme Trust for the award of a Research Fellowship. Finally, we wish thank Fran\c cois Mignard for providing the Famous software.
We wish to thank Joseph N. Spitale and the two others reviewers for their careful reading of our paper and many excellent suggestions.

\section{Appendix A}
\begin{table}[H]
\tiny
\center
\begin{tabular}{|l|c|r|r|r|r|} 
   \hline
     Date (Year-day)  & Amplitudes (km) & Phase (deg)  &  m    \\
    \hline
    \hline  
    2006 - 271 &  10.40    &    50.50  &      5.78    \\
    \cline{2-4} 
       		     &  6.88    &    8.67   &      3.78    \\
    \cline{2-4}
        		     & 4.98  &      309.95   &   8.72     \\
     \hline  
     \hline   
     2006 - 304    & 7.73  &       64.71  &        5.47    \\
    \cline{2-4} 
        		         &  6.30  &      190.83  &      6.37     \\
     \cline{2-4}    
                         & 6.84  &      160.34 &       4.39  \\ 
     \hline 
     \hline 
     	2006 - 316   & 10.17  &      255.72 &       6.00  \\ 
    \cline{2-4} 
        		        &  4.75   &      210.86  &       4.03  \\  
     \cline{2-4}     
                        & 5.20  &      218.06 &       7.02   \\ 
     \hline 
     \hline 
     2006 - 329   &   7.14  &        26.23  &        3.82  \\ 
    \cline{2-4} 
        		       & 4.52  &      172.53  &       9.21  \\  
     \cline{2-4}     
                        &   4.08  &       130.02 &         1.71  \\    
\hline 
     \hline 
     2006 - 357   & 5.90  &        272.60  &        4.01  \\ 
    \cline{2-4} 
        		        & 4.51  &        296.43  &        1.92   \\  
     \cline{2-4}     
                        &   4.33  &       321.07  &        7.72  \\   
     \hline 
  \hline 
   2007 - 005     &  10.89  &       235.93  &        6.02  \\ 
    \cline{2-4} 
        		        &  6.29  &       205.13  &       4.06  \\  
     \cline{2-4}     
                        &  5.93  &        244.38  &       7.50   \\    
   \hline 
  \hline 
    2007 - 041   & 12.40   &      132.00  &        5.97   \\ 
    \cline{2-4} 
        		        &  8.50  &        134.37  &        3.87   \\  
     \cline{2-4}     
                        & 4.74   &       253.12  &        1.76    \\    
   \hline 
  \hline 
     2007 - 058   &  12.56  &       40.09   &        5.87   \\ 
    \cline{2-4} 
        		        & 6.14     &      332.05    &      6.89    \\  
     \cline{2-4}     
                        &  5.43     &       344.88   &     4.05   \\    
   \hline 
  \hline 
   2007 - 076     & 10.90     &       334.03    &     6.18    \\ 
    \cline{2-4} 
        		        & 8.18      &   280.13      &  4.09   \\  
     \cline{2-4}     
                        &   1.76    &      288.27    &      2.02    \\    
   \hline 
  \hline 
  2007 - 090     &  17.93    &       255.90     &     6.16  \\ 
    \cline{2-4} 
        		       &  8.41     &    252.28    &    3.69 \\  
     \cline{2-4}     
                        &  1.86     &    52.30    &     8.81  \\    
   \hline 
  \hline 
 2007 - 125  &    21.56   &    168.88   &     6.21  \\ 
    \cline{2-4} 
        		     &  9.46     &   359.54   &     6.87  \\  
     \cline{2-4}     
                      &  7.77    &     88.60   &      4.08  \\    
   \hline 
     \hline    
2008 - 023    & 17.71 &      334.47   &        5.79    \\ 
    \cline{2-4} 
        		       & 9.25   &     13.83   &      4.52   \\  
     \cline{2-4}     
                      &   7.60     &   267.75     &   7.48  \\    
   \hline 
  \hline      
    2008 - 243    &  14.33 &       252.42  &      4.30   \\ 
    \cline{2-4} 
        		        &  13.42   &      323.89    &       6.34   \\  
     \cline{2-4}     
                       &  6.54  &       23.43  &       9.39 \\    
   \hline 
  \hline 
 2008 - 274      & 11.00     &   35.17   &     6.05   \\ 
    \cline{2-4} 
        		        & 7.14     &   285.54   &     3.69  \\  
     \cline{2-4}     
                       & 4.85      &   24.32    &    8.57  \\    
   \hline    
     \hline  
 2008 - 288     &  14.24    &     0.81     &    5.87 \\ 
    \cline{2-4} 
        		       &  7.48    &     283.77    &   3.94 \\  
     \cline{2-4}     
                       &  4.95     &    22.65    &     9.40  \\    
   \hline 
  \hline    
     2008 - 303      &  13.77   &       228.94   &      5.92   \\ 
    \cline{2-4} 
        		      &  7.36     &    97.06    &    3.94 \\  
   \hline 
  \hline   
   2009 - 011   &  13.93  &      295.00   &      6.05   \\ 
    \cline{2-4} 
        		     &  5.96    &      154.45   &      8.59  \\  
     \cline{2-4}     
                     &  5.75   &       147.04    &     3.76 \\    
   \hline 
  \hline 
    2009 - 041        &  16.75     &   151.71      &     5.58   \\ 
    \cline{2-4} 
        		       &   4.13     &      280.71    &      3.26   \\  
   \hline 
   \hline    
    2009 - 070    &  11.66    &     121.57   &        5.81  \\ 
    \cline{2-4} 
        		      &  4.37    &      324.17  &        3.37  \\  
     \cline{2-4}     
                      &   3.33    &       329.46   &       7.38   \\    
   \hline 
  \hline 
2009 - 082     &   11.84    &       102.25    &    6.02  \\ 
    \cline{2-4} 
        		      &  2.15    &         139.45      &  2.01   \\  
   \hline 
  \hline 
  2009 - 106    &  10.49    &    168.88    &     6.38  \\ 
    \cline{2-4} 
        		      &   8.54   &       330.29  &       3.74 \\  
     \cline{2-4}     
                       &  6.23   &       288.85   &       4.76 \\    
   \hline 
  \hline    
    2009 - 130      &   8.45      &    11.40        &   4.22  \\ 
   \hline 
  \hline
  2009 - 211      &   11.78   &      136.43    &    5.97   \\ 
    \cline{2-4} 
        		      &  4.56   &         213.76   &      2.26 \\     
   \hline             
\end{tabular}
\caption{Frequency analysis results for imaging mosaics obtained between 2006 and 2009, in the local frame $\Omega^{*} = n = 604.22\dd$.}
\label{table_famous_edge_before}
\end{table} 
    \begin{table}[H]
    \tiny
    \center
\begin{tabular}{|l|c|r|r|r|} 
   \hline
     Date (Year-day)  & Amplitudes (km) & Phase (deg)  &  m    \\
   \hline 
   \hline 
 2013 - 077     &   18.83     &  203.53    &     6.40  \\ 
    \cline{2-4} 
        		      &   9.52     &   141.48   &        3.89   \\  
     \cline{2-4}     
                      &  3.92   &     203.50   &     9.22  \\    
   \hline 
  \hline  
    2013 - 126     & 8.86   &        251.15   &      4.09   \\ 
    \cline{2-4} 
        		       &   8.36       &  218.384   &    2.31 \\  
     \cline{2-4}     
                        &  5.53     &    215.87     &    8.40 \\    
   \hline 
  \hline 
 2013 - 147      &  8.17    &       187.58     &     3.90   \\ 
    \cline{2-4} 
        		       &  6.52   &      303.53  &        6.46  \\  
     \cline{2-4}     
                        &  4.48    &      235.84   &       2.65  \\    
   \hline 
  \hline   
  2013 - 232      &  11.68   &     236.33     &    3.72 \\ 
    \cline{2-4} 
        		        &  7.60    &       219.10    &       1.68  \\  
     \cline{2-4}     
                        &  3.99    &      266.24    &     9.10  \\    
   \hline               
   \hline 
   2013 - 236       &  9.35     &     251.67    &       3.94    \\ 
    \cline{2-4} 
        		        & 6.91     &     240.81        &    1.78   \\  
     \cline{2-4}     
                        & 2.87   &      302.96     &    8.97    \\    
   \hline 
  \hline 
    2013 - 250       &  16.95     &    290.09    &    4.45 \\  
   \hline 
  \hline   
  2013 - 291      &   9.14    &     321.11  &        6.50   \\ 
    \cline{2-4} 
        		        &  5.94    &     226.82    &     8.47 \\  
     \cline{2-4}     
                        &  4.30     &     169.81     &    1.57  \\    
   \hline 
  \hline 
  2014 - 103     &   4.88    &      10.29    &      4.44   \\ 
    \cline{2-4} 
        		        &  3.01   &      41.53     &     6.45  \\     
   \hline 
  \hline 
 2014 - 173    &  4.72   &      176.74    &     4.12     \\ 
    \cline{2-4} 
        		        &  4.52   &      230.03   &       6.08  \\  
   \hline  
\end{tabular}
\caption{Frequency analysis results for imaging mosaics obtained between 2013 and 2014, in the local frame $\Omega^{*} = n = 604.22\dd$.}
\label{table_famous_edge_after}
\end{table}

\newpage

\includepdf[pages={1-1}]{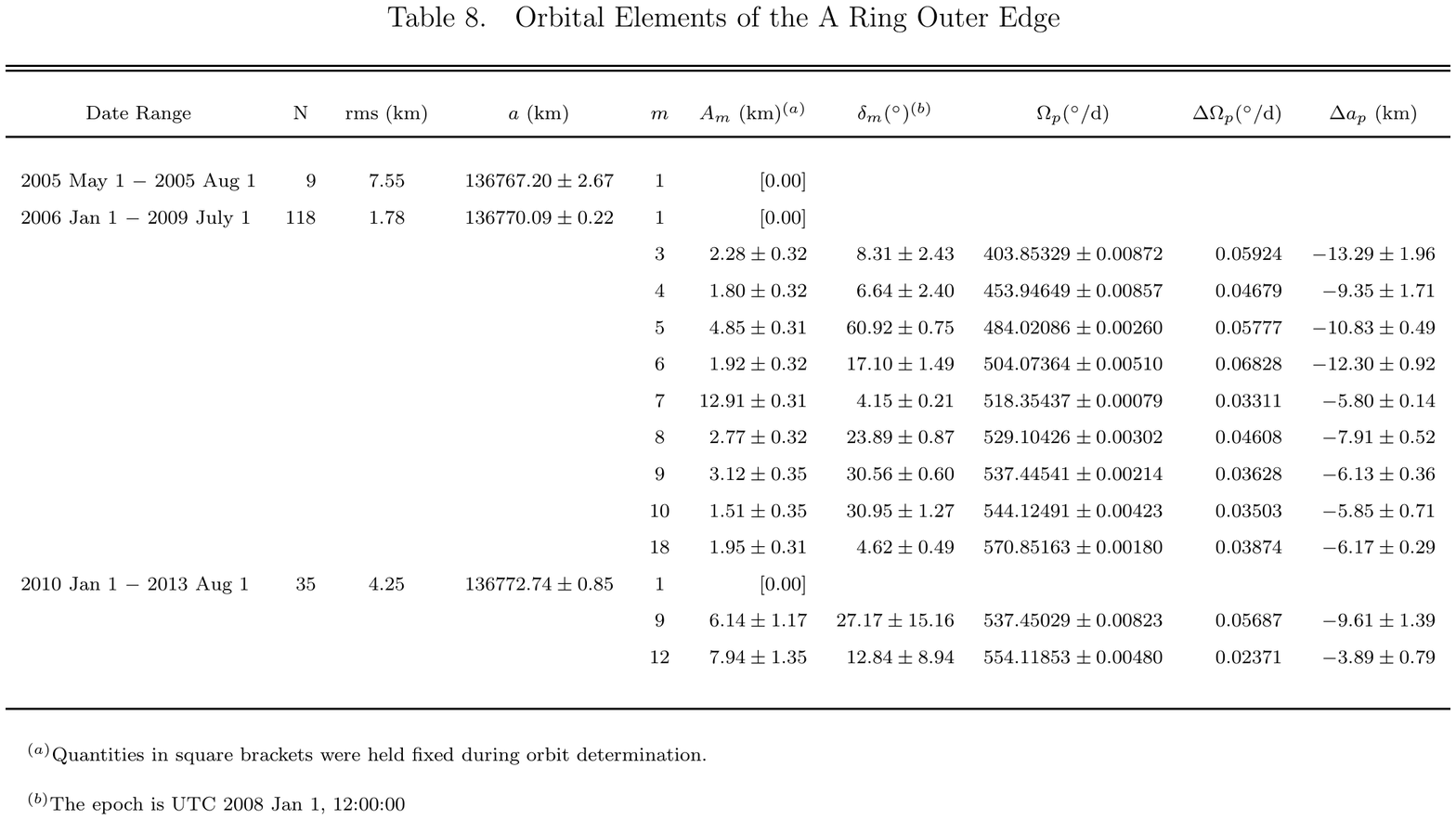}


\footnotesize
\nocite{*}
\bibliographystyle{icarus}    
\bibliography{mnemonic,biblio}   

\if{

}\fi

\end{document}